\documentclass[largeformat]{interact}

\usepackage{epstopdf}
\usepackage[caption=false]{subfig}

\usepackage[numbers,sort&compress]{natbib}
\bibpunct[, ]{[}{]}{,}{n}{,}{,}
\makeatletter
\def\NAT@def@citea{\def\@citea{\NAT@separator}}
\makeatother

\theoremstyle{plain}

\theoremstyle{definition}

\theoremstyle{remark}

\usepackage{amssymb, amsmath, amsthm, mathtools}
\usepackage{bm}
\usepackage{mathrsfs}
\usepackage{hyperref,bookmark}
\usepackage{graphicx}
\usepackage{epsfig,color}
\usepackage{mathrsfs}
\usepackage{verbatim}
\usepackage{url}
\usepackage{float}
\usepackage[greek,english]{babel}
\usepackage{dcolumn}

\newcommand{\tr}{\operatorname{tr}}
\newcommand{\dd}{\operatorname{d}\!}

\newcommand{\diver}{\operatorname{div}}
\newcommand{\curl}{\operatorname{curl}}

\newcommand{\n}{\bm{n}}
\newcommand{\e}{\bm{e}}

\newcommand{\normal}{\bm{\nu}}
\newcommand{\body}{\mathscr{B}}
\newcommand{\surface}{\mathscr{S}}
\newcommand{\free}{\mathscr{F}}

\newcommand{\vt}{\vartheta}
\newcommand{\vp}{\varphi}

\newcommand{\Q}{\mathbf{Q}}
\newcommand{\R}{\mathbf{R}}

\newcommand{\nablas}{\nabla\!_\mathrm{s}}

\newcommand{\trans}{^\mathsf{T}}

\newcommand{\WOF}{W_\mathrm{OF}}
\newcommand{\WQT}{W_\mathrm{QT}}
\newcommand{\bend}{\bm{b}}
\newcommand{\Wn}{\mathbf{W}(\n)}
\newcommand{\W}{\mathbf{W}}
\newcommand{\Wz}{\W_z}
\newcommand{\Pn}{\mathbf{P}(\n)}
\renewcommand{\Pr}{\mathbf{P}_r}
\newcommand{\Dn}{\mathbf{D}}
\newcommand{\I}{\mathbf{I}}
\newcommand{\twon}{(\n_1,\n_2)}
\newcommand{\zero}{\bm{0}}

\newcommand{\nT}{\n_{\mathrm{T}}}
\newcommand{\nR}{\n_{\mathrm{R}}}

\newcommand{\nH}{\n_{\mathrm{H}}}
\newcommand{\vv}{\bm{v}}
\newcommand{\xv}{\bm{x}}

\newcommand{\et}{\e_\vt}
\newcommand{\ep}{\e_\vp}
\newcommand{\degree}{^\circ}
\newcommand{\ballx}{\mathbb{B}_R(\bm{x}_0)}
\newcommand{\ball}{\mathbb{B}}
\newcommand{\cyl}{\mathbb{C}}
\newcommand{\sphere}{\mathbb{S}^2}

\newcommand{\parity}[1]{\overline{#1}}

\newcommand{\act}{\mathcal{A}}

\begin{document}

\title{Spiraling Defect Cores in Chromonic Hedgehogs}

\author{
	\name{Silvia Paparini\textsuperscript{a} and Epifanio G. Virga\textsuperscript{b}\thanks{CONTACT Epifanio G. Virga. Email: eg.virga@unipv.it}}
	\affil{\textsuperscript{a}\textsuperscript{b}Dipartimento di Matematica, Universit\`a di Pavia, Via Ferrata 5, 27100 Pavia, Italy}
}

\maketitle

\begin{abstract}
An elastic \emph{quartic twist theory} has recently been proposed for chromonic liquid crystals, intended to overcome the paradoxical conclusions encountered by the classical Oseen-Frank theory when applied to droplets submerged in an isotropic fluid environment. However, available  experimental data for chromonics confined to cylindrical cavities with degenerate planar anchoring on their lateral boundary can be explained equally well by both competing theories. This paper identifies a means to differentiate these theories both qualitatively and quantitatively. They are shown to predict quite different core defects for the \emph{twisted} hedgehogs that chromonics generate when confined to a fixed spherical cavity with homeotropic anchoring. In the quartic twist theory, the defect core is estimated to be nearly one order of magnitude larger (tens of microns) than in the other and, correspondingly, the director field lines describe Archimedean spirals instead of logarithmic ones.
\end{abstract}



\section{Introduction}\label{sec:intro}
The classical theory of nematic curvature elasticity is based on the assumption that in the ground state the director $\n$ has everywhere the same orientation; the energy stored in a distortion then measures the work done to produce it starting from the ground state. The classical Oseen-Frank theory posits a stored energy quadratic in $\nabla\n$ and features four elastic constants, one for each elementary distortional mode.

Chromonic liquid crystals (CLCs) are lyotropic materials, which include Sunset Yellow (SSY), a popular dye in food industry, and disodium cromoglycate (DSCG), an anti-asthmatic drug. In these materials, molecules stuck themselves in columns, which in aqueous solutions develop a nematic orientational order. In CLCs, $\n$ designates the average direction in space of the constituting  supra-molecular aggregates. 

A number of reviews have progressively become available in the last few years \cite{lydon:chromonic_1998,lydon:handbook,lydon:chromonic_2010,lydon:chromonic,dierking:novel}; we refer the interested reader to them. 

Experiments have been performed with these materials in capillary tubes, with either circular \cite{nayani:spontaneous,davidson:chiral} or rectangular \cite{fu:spontaneous} cross-sections, as well as on cylindrical shells \cite{javadi:cylindrical}, all enforcing \emph{degenerate planar} anchoring, which allows constituting columns to glide freely on the anchoring surface, provided they remain tangent to it. These experiments revealed a tendency of CLCs to acquire spontaneously a \emph{double twist} configuration in cylinders. Due to the lack of chirality in the molecular aggregates constituting CLCs, spontaneous double twists come equally likely in two variants with opposite chiralities. 

Despite the lack of \emph{uniformity} in the ground state of these phases,\footnote{The  classification of the most general \emph{uniform} distortions, which can fill the whole three-dimensional space, is given in \cite{virga:uniform} and recalled in Sect.~\ref{sec:quadratic_formula}.} their curvature elasticity has been modeled by the  Oseen-Frank theory, albeit with an anomalously small twist constant $K_{22}$. To accommodate the experimental findings and justify the (double) twisted ground state, this constant has to be smaller than the saddle-splay constant $K_{24}$, in violation of one of the inequalities Ericksen~\cite{ericksen:inequalities} had put forward to guarantee that the Oseen-Frank stored energy be bounded below.

Actually, as shown in \cite{paparini:stability}, the violation of one Ericksen's inequality does not prevent the twisted ground state from being locally stable in a cylinder enforcing degenerate planar anchoring on its lateral boundary. The same  conclusion was reached in \cite{long:violation} on different grounds. But, as shown in \cite{paparini:paradoxes}, free-boundary problems may reveal noxious consequences of violating Ericksen's inequalities.   If $K_{22}<K_{24}$, a CLC droplet, tactoidal\footnote{\emph{Tactoids} are elongated, cylindrically symmetric shapes with pointed ends as poles.} in shape and surrounded by an isotropic fluid environment enforcing degenerate planar anchoring for the director at the interface, is predicted to be unstable against \emph{shape} perturbations: it would split indefinitely in smaller tactoids while the total free energy plummets to negative infinity (see \cite{paparini:paradoxes}, for more details).

This prediction is in sharp contrast with the wealth of experimental observations of CLC tactoidal droplets, stable in the biphasic region of phase space, where nematic and isotropic phases coexist in equilibrium. Experiments have been carried out with a number of substances (including DSCG and SSY) stabilized by the addition of neutral (achiral) condensing agents (such as PEG and Spm) \cite{tortora:self-assembly,tortora:chiral,peng:chirality,nayani:using,shadpour:amplification}. These studies have consistently reported stable twisted bipolar tactoids.
	
To resolve this contradiction, in \cite{paparini:elastic} we proposed a minimalist quartic theory for CLCs, which adds to the Oseen-Frank energy density a single quartic term in the (double) twist measure; hence the name \emph{quartic twist} theory. We showed in \cite{paparini:elastic} that indeed within this theory the total free energy of chromonic droplets subject to degenerate planar interfacial anchoring remains bounded below, even if $K_{22}<K_{24}$; we also used published data to prove consistency with experiments and estimated a phenomenological length introduced by the theory.  

Higher-order theories are not new in liquid crystal science. Go under this name either theories that allow for higher spatial gradients of $\n$ in the energy and theories that allow for higher powers in the first gradient. Under the first category, which perhaps has seen its first manifestation in \cite{nehring:elastic} (see also \cite{oldano:ab}), falls, for example, Dozov's theory \cite{dozov:spontaneous} for both twist-bend and splay-bend phases predicted long ago by Meyer \cite{meyer:structural} and more recently observed in real materials \cite{cestari:phase}. Under the second category falls, for example, a simple one-dimensional model for splay-bend nematics \cite{lelidis:nematic}, then extended to incorporate a whole class of seven modulated ground states, of which twist-bend and splay-bend are just two instances \cite{barbero:fourth}.
A hybrid theory was also proposed in \cite{lelidis:nonlinear}, where both higher gradients of $\n$ and higher powers of the first gradient are allowed in the stored-energy density, with spatial derivatives and their powers balanced according to a criterion motivated by a molecular model.\footnote{It should also be noted that other theories are known as ``quartic'' (see, for, example, the classical paper \cite{longa:extension} and the more recent contribution \cite{golovaty:novel}), but they owe this name to an elastic term globally quartic in de~Gennes' order tensor and its derivatives, added to the commonly considered version of the Landau de Gennes theory to resolve the spay-bend elastic constant degeneracy in the reduction to the Oseen-Frank theory. These theories serve a different purpose.} Our quartic theory is much simpler than these.

In \cite{paparini:elastic}, we showed that both the classical Oseen-Frank theory and our quartic twist theory explain  experimental data for the emergence of double twist in capillaries to a comparable degree of confidence. Here, in our quest for qualitative and quantitative features that may allow us to discriminate between these theories, we consider the case of the most common of point defects, the \emph{hedgehog}.

We imagine a CLC confined within a fixed spherical cavity enforcing homeotropic anchoring on its boundary, as in a recent experiment \cite{spina:intercalation}. Since the seminal paper of Lavrentovich and Terentjev~\cite{lavrentovich:phase} we know that for $K_{22}$ sufficiently small a radial hedgehog becomes \emph{twisted} and exhibits field lines spiraling about the point defect. A defect \emph{core} can then be easily identified; geometrically, it is delimited by an \emph{inversion ring} where spirals invert their winding sense. This is the feature under close scrutiny here: We want to mark the differences between quadratic and quartic theories in describing the defect core of a twisted hedgehog. We are interested in qualitative and quantitative differences as well, aiming to outline a setting that could possibly discriminate one theory from the other.

The paper is organized as follows. In Sect.~\ref{sec:energetics}, we describe the energetics of chromonics, starting from the classical quadratic Oseen-Frank theory and then summarizing our quartic twist theory. In Sect.~\ref{sec:twisted_hedg}, we describe a class of director fields intended to represent twisted hedgehogs in a ball in terms of a single \emph{twist} angle depending on the radial coordinate only. We review the conditions that ensure that the radial hedgehog is unstable (according to both elastic theories) and find the energy-minimizing twist angle for the quartic twist theory; for the quadratic theory, this problem was solved in \cite{ball:brief}. The results for the two theories are then compared and their stark differences emerge in Sect.~\ref{sec:cores}. Finally, in Sect.~\ref{sec:discussion}, we summarize the conclusions of our work and comment on possible avenues for future research.

The paper is closed by two technical appendices. In one, we collect a number of mathematical details concerning our representation of twisted hedgehogs. In the other, we describe an equivalent dynamical system, whose orbits correspond to twisted hedgehogs in equilibrium. A similar correspondence was used in \cite{ball:brief}, with the further advantage (lost here) that the equivalent dynamical system was autonomous.
  
\section{Quadratic and Quartic Theories}\label{sec:energetics}
As customary in liquid crystal science, an elastic theory for chromonics is based on a free-energy functional $\free$ that expresses the energy stored in a region in space $\body$  containing the material as
\begin{equation}
	\label{eq:free_energy_functional}
	\free[\n]:=\int_{\body}W(\n,\nabla\n)\dd V,
\end{equation}
where $W$ is a function of the nematic director $\n$ and its gradient $\nabla\n$, which here plays the role of a local (tensorial) measure of distortion, and $\dd V$ is the volume element.

We start by summarizing the classical quadratic theory for the elasticity of nematic liquid crystals, albeit formulated in a novel, equivalent way that serves better our purpose. It will then become easier to present the quartic twist theory proposed in \cite{paparini:paradoxes}. 

\subsection{Classical Quadratic Energy}\label{sec:quadratic_formula}
The classical elastic theory of liquid crystals goes back to the pioneering works of Oseen~\cite{oseen:theory} and Frank~\cite{frank:theory}.\footnote{Also a paper by Zocher~\cite{zocher:effect}, mainly concerned with the effect of a magnetic field on director distortions, is often mentioned among the founding contributions. Some authors go to the extent of also naming the theory after him. Others, in contrast, name the theory only after Frank, as they only deem his contribution to be fully aware of the nature of $\n$ as a \emph{mesoscopic}  descriptor of molecular order.} In this theory, the elastic free-energy density $W$ in \eqref{eq:free_energy_functional} is chosen to be the most general frame-indifferent,\footnote{This requirement amounts to assume that $W(\Q\n,\Q(\nabla\n)\Q\trans)=W(\n,\nabla\n)$, for all rotations $\Q$ in three-dimensional space.}  even function quadratic in $\nabla\n$, 
\begin{equation}
	\label{eq:free_energy_density}
	\begin{split}
	W=\WOF(\n,\nabla\n)&:=\frac{1}{2}K_{11}\left(\diver\n\right)^2+\frac{1}{2}K_{22}\left(\n\cdot\curl\n\right)^2+ \frac{1}{2}K_{33}|\n\times\curl\n|^{2}\\ &+ K_{24}\left[\tr(\nabla\n)^{2}-(\diver\n)^{2}\right].
	\end{split}
\end{equation}
Here $K_{11}$, $K_{22}$, $K_{33}$, and $K_{24}$ are elastic constants characteristic of the material. They are traditionally referred to as the \emph{splay}, \emph{twist}, \emph{bend}, and \emph{saddle-splay} constants, respectively, by the features of four different orientation fields, each with a distortion energy proportional to a single term in \eqref{eq:free_energy_density} (see, for example, Chap.~3 of \cite{virga:variational}). 

Recently, Selinger~\cite{selinger:interpretation} has reinterpreted the classical formula \eqref{eq:free_energy_density} by decomposing the saddle-splay mode into a set of other independent modes. The starting point of this decomposition is a novel representation of $\nabla\n$ (see also \cite{machon:umbilic}),
\begin{equation}
	\label{eq:nabla_n_novel}
	\nabla\n=-\bend\otimes\n+\frac12T\Wn+\frac12S\Pn+\Dn,
\end{equation}
where $\bend:=-(\nabla\n)\n=\n\times\curl\n$ is the \emph{bend} vector, $T:=\n\cdot\curl\n$ is the \emph{twist}, $S:=\diver\n$ is the \emph{splay}, $\Wn$ is the skew-symmetric tensor that has $\n$ as axial vector, $\Pn:=\I-\n\otimes\n$ is the projection onto the plane orthogonal to $\n$, and $\Dn$ is a symmetric tensor such that $\Dn\n=\zero$ and $\tr\Dn=0$. By its own definition, $\Dn\neq\zero$ admits the following biaxial representation,
\begin{equation}
	\label{eq:D_representation}
	\Dn=q(\n_1\otimes\n_1-\n_2\otimes\n_2),
\end{equation}
where $q>0$ and $\twon$ is a pair of orthogonal unit vectors in the plane orthogonal to $\n$, oriented so that $\n=\n_1\times\n_2$.\footnote{It is argued in \cite{selinger:director} that $q$ should be given the name \emph{tetrahedral} splay, to which we would actually prefer \emph{octupolar} splay for the role played by a cubic (octupolar) potential on the unit sphere \cite{pedrini:liquid} in representing all scalar measures of distortion,  but $T$.} In the local frame $(\n_1,\n_2,\n)$, $\bend$ is represented as
\begin{equation}
	\label{eq:b_1_b_2}
	\bend=b_1\n_1+b_2\n_2.
\end{equation}

By use of the following identity, 
\begin{equation}
	\label{eq:identity}
	2q^2=\tr(\nabla\n)^2+\frac12T^2-\frac12S^2,
\end{equation}
we can easily give \eqref{eq:free_energy_density} the equivalent form
\begin{equation}
	\label{eq:Frank_equivalent}
	\WOF(\n,\nabla\n)=\frac12(K_{11}-K_{24})S^2+\frac12(K_{22}-K_{24})T^2+\frac12K_{33}B^2+2K_{24}q^2,
\end{equation}
where $B^2:=\bend\cdot\bend=b_1^2+b_2^2$. Since $(S,T,b_1,b_2,q)$ are all independent \emph{distortion characteristics}, it readily follows from \eqref{eq:Frank_equivalent} that $\WOF$ is positive semi-definite whenever
\begin{subequations}\label{eq:Ericksen_inequalities}
	\begin{eqnarray}
		&K_{11}\geqq K_{24}\geqq0,\label{eq:Ericksen_inequalities_1}\\
		&K_{22}\geqq K_{24}\geqq0, \label{eq:Ericksen_inequalities_2}\\
		&K_{33} \geqq 0,\label{eq:Ericksen_inequalities_3}
	\end{eqnarray}
\end{subequations}
which are the celebrated \emph{Ericksen's inequalities} \cite{ericksen:inequalities}. If these inequalities are satisfied in strict form, the global ground state of $\WOF$ is attained on the uniform director field, characterized by
\begin{equation}
	\label{eq:uniform_ground_state}
	S=T=B=q=0.
\end{equation}
As already mentioned in the Introduction, inequality \eqref{eq:Ericksen_inequalities_2} must be violated for the ground state of $\WOF$ to be different from \eqref{eq:uniform_ground_state}, involving a non-vanishing $T$. 

The class of \emph{uniform} distortions was defined in \cite{virga:uniform} as the one comprising all director fields for which the distortion characteristics are constant in space. Equivalently said, a uniform distortion is a director field that can \emph{fill} three-dimensional space. It was proven that there are two distinct families of uniform distortions, characterized by the following conditions \cite{virga:uniform},
\begin{equation}\label{eq:uniform_distortions}
S=0,\quad T=\pm2q,\quad b_1=\pm b_2=b,	
\end{equation} 
where $q$ and $b$ are arbitrary parameters.

The general director field corresponding to \eqref{eq:uniform_distortions} is the \emph{heliconical} ground state of twist-bend nematic phases,\footnote{With opposite \emph{chiralities}, one for each sign in \eqref{eq:uniform_distortions}.} in which $\n$ makes a fixed \emph{cone} angle with a given axis in space (called the \emph{helix} axis), around which $\n$ precesses periodically \cite{virga:uniform}.\footnote{In opposite senses, according to the sign of chirality.} The special instance in which $b=0$ corresponds to the \emph{single twist} that characterizes \emph{cholesteric} liquid crystals. 

The distortion for which all characteristics vanish, but $T$, is a \emph{double twist}.\footnote{Here we adopt the terminology proposed by Selinger~\cite{selinger:director} (see also \cite{long:explicit}) and distinguish between \emph{single} and \emph{double} twists, the former being \emph{uniform} and the latter not.} It is \emph{not} uniform and cannot fill space; it can possibly be realized locally, but not everywhere. In words, we say that it is a \emph{frustrated} ground state. As shown in \cite{paparini:stability}, a double twist is indeed attained exactly only on the symmetry axis of cylinders enforcing degenerate planar anchoring on their lateral boundary. 

\subsection{Quartic Twist Energy}\label{sec:quartic_formula}
The essential feature of the quartic twist theory proposed in \cite{paparini:elastic} is to envision a double twist with two equivalent chiral variants as ground state of CLCs in three-dimensional space,
\begin{equation}
\label{eq:double_twist}
S = 0, \quad T = \pm T_0, \quad B = 0, \quad q = 0.
\end{equation}
The degeneracy of the ground  double twist  in \eqref{eq:double_twist} arises from  the achiral nature of the molecular aggregates that constitute these materials, which is reflected in the lack of chirality of their condensed phases.

The elastic stored energy must equally penalize both ground chiral variants. Our minimalist  proposal to achieve this goal was to add  a \emph{quartic twist} term to the Oseen-Frank stored-energy density, and so take $W=\WQT$, with
\begin{equation}
\label{eq:quartic_free_energy_density}
\WQT(\n,\nabla\n):=\frac{1}{2}(K_{11}-K_{24})S^2+\frac{1}{2}(K_{22}-K_{24})T^2+ \frac{1}{2}K_{23}B^{2}+\frac{1}{2}K_{24}(2q)^2 + \frac14K_{22}a^2T^4,
\end{equation}
where $a$ is a \emph{characteristic length}. Unlike $\WOF$, $\WQT$ is bounded below whenever 
\begin{subequations}\label{eq:new_inequalities}
\begin{eqnarray}
	&K_{11}\geqq K_{24}\geqq0,\label{eq:new_inequalities_1}\\
&K_{24}\geqq K_{22}\geqq0, \label{eq:new_inequalities_2}\\
	&K_{33}\geqq0.\label{eq:new_inequalities_3}
\end{eqnarray}
\end{subequations}
If these inequalities  hold, as we shall assume here, then $\WQT$ is minimum at the  degenerate double-twist \eqref{eq:double_twist}
characterized by
\begin{equation}
	\label{eq:T_0min}
	T_0:=\frac{1}{a}\sqrt{\frac{K_{24}-K_{22}}{K_{22}}}.
\end{equation}
The parameter $a$ encodes the \emph{bare} length scale over which distortions would be locally stored in the ground state.\footnote{In the elastic model proposed in \cite{virga:uniform} for twist-bend nematics, a quartic free energy was posited that admits as ground state either of two families of uniform heliconical fields with opposite chirality. There too, a length scale appears in the equilibrium pitch. The distortion state characterized by this  length is the same everywhere.} As to the physical size of such a length scale, it may be comprised in a wide range. While at the lower end we may place the persistence length of the molecular order, which characterizes the flexibility of CLC aggregates,\footnote{The persistence length of a flexible aggregate is the shortest length over which unit vectors tangent to the aggregate's contour lose correlation. For CLCs, it is estimated on the order of tens to hundreds of $\mathrm{nm}$ \cite{zhou:lyotropic}} the upper end is hard to make definite. We expect that $a$ would be exposed to the same indeterminacy that affects many (if not all) supramolecular structures in lyotropic systems. The most telling example is perhaps given by cholestric liquid crystals, which give rise to a chiral structure (characterized by a single twist $T=\pm2q$) starting from chiral molecules. If the macroscopic pitch were determined by the molecular chirality,\footnote{\emph{Via} the naive geometric argument that represents chiral molecules as cylindrical screws and derives the pitch of their assemblies by close packing them so as to fit grooves with grooves.} it would result several orders of magnitude smaller than the observed ones.\footnote{For lyotropic cholesterics, the mismatch between microscopic and macroscopic pitches, which has recently received new experimental evidence in  systems of basic living constituents \cite{stanley:dna,tortora:chiral_2011}, is still debated. Interesting theories based on either molecular shape fluctuations \cite{harris:microscopic,harris:molecular} or surface charge patterns \cite{kornyshev:chiral} have met with some experimental disagreement \cite{grelet:what}.} Here, we shall treat $a$ as a phenomenological parameter, to be determined experimentally.
An estimate derived in \cite{paparini:elastic} from a comparison with published data placed $a$ in the order of microns.

\section{Twisted Hedgehog}\label{sec:twisted_hedg}
So far we have presented, mostly on equal terms, two elastic theories for chromonics, one quadratic and the other quartic in the director gradient. Here we see how these theories can be differentiated on the basis of the different structures they predict for the core of hedgehogs, the most common of nematic defects in three space dimensions. Many mathematical details needed to follow our development are collected in Appendix~\ref{sec:trial_twisted}.

We first discuss the distortion of a trial director field within a ball of radius $R$ enforcing homeotropic anchoring on its boundary. This is a field with a point defect at the center of the ball, potentially rich in twist, as would seem  fit for a material with small $K_{22}$ constant. We shall then see the analytical implications and the potential experimental significance of this field. 

\begin{figure}
	\centering
	\subfloat[Radial hedgehog: $N(\nR)=+1$]{%
		\resizebox*{5.5cm}{!}{\includegraphics{n_R.pdf}}}\hspace{50pt}
	\subfloat[Hyperbolic hedgehog: $N(\nH)=+1$]{%
		\resizebox*{5.5cm}{!}{\includegraphics{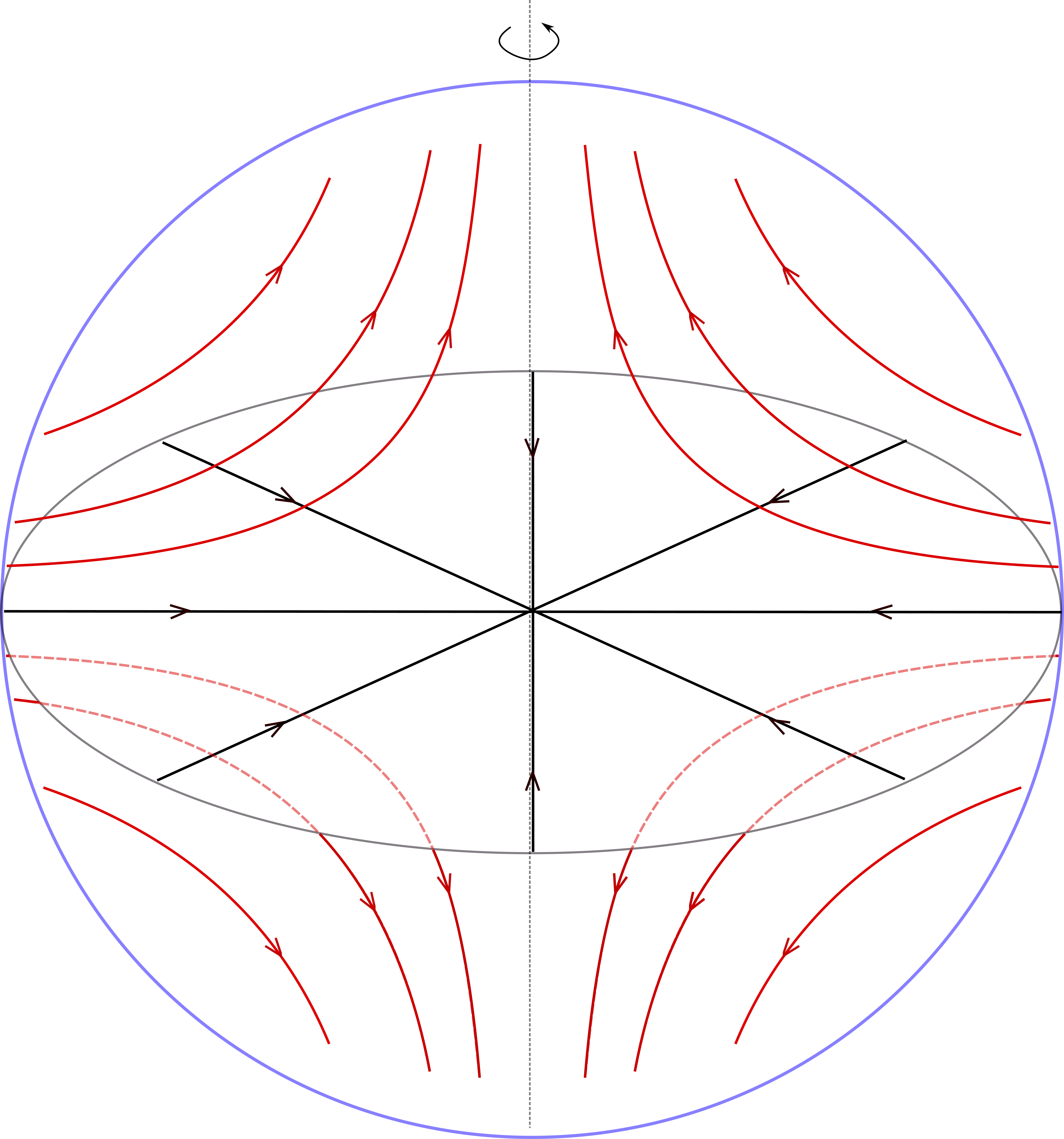}}}
	\caption{Field lines of $\nR$ and $\nH$ in \eqref{eq:rad_hedg} and \eqref{eq:hip_hedg}, representing a radial and a hyperbolic hedgehog, respectively. Field lines are drawn on the equatorial plane (in black) and on a meridian plane (in red). The whole picture is obtained by rotating these lines around the polar axis. By their definitions, both $\nR$ and $\nH$ share the same topological charge $N$ as introduced in \eqref{eq:topological_charge}.}
	\label{fig:radial_hyperbolic_hedgehogs}
\end{figure}

The point defects that we shall study are a special family of \emph{hedgehogs}, which place themselves in between the most common defects in liquid crystal science, the \emph{radial} and the \emph{hyperbolic} hedgehogs. The former is represented by the director field
\begin{equation}
	\label{eq:rad_hedg}
	\nR(\xv):=\frac{\xv-\bm{x}_0}{|\xv-\bm{x}_0|},
\end{equation}
which has a point defect at $\xv_0$, while the latter is formally obtained  by the following transformation of $\nR$,
\begin{equation}
	\label{eq:hip_hedg}
	\nH:=\R(\pi)\nR,
\end{equation}
where
\begin{equation}
	\label{eq:rotation_pi}
	\R(\pi):=-\I+2\e\otimes\e
\end{equation}
is the special orthogonal tensor  describing a rotation by angle $\pi$ about a unit vector $\e\in\mathbb{S}^2$.  Figure~\ref{fig:radial_hyperbolic_hedgehogs} illustrates  the field lines of both $\nR$ and $\nH$.

The \emph{topological charge} of a unit vector field $\n$ with a point defect at $\xv_0$ is defined as
\begin{equation}
	\label{eq:topological_charge}
	N(\n):=\frac{1}{4\pi}\int_{\surface}\n\cdot(\nablas\n)^*\normal\dd A,
\end{equation}
where $\surface$ is a any surface enclosing $\xv_0$, $\nablas$ denotes the surface gradient on $\surface$, $\normal$ is the unit normal to $\surface$, the operation $(\cdots)^*$ takes the cofactor of a tensor,\footnote{This is a tensor whose representative matrix is the cofactor matrix of the matrix representing the original tensor, see \cite[p.\,22]{gurtin:mechanics} for a formal definition.} and $\dd A$ is the area element. $N(\n)$ is an integer of $\mathbb{Z}$ independent of $\surface$, provided the latter embraces $\xv_0$, and so $N(\n)$ can be attributed to $\xv_0$ itself. The absolute value $|N(\n)|$ indicates the number of times $\n$ restricted to $\surface$ covers the unit sphere $\sphere$; the sign of $N(\n)$ tells whether $\sphere$ is covered coherently or not with the orientation of the unit normal $\normal$. Historically, we learn from \cite{kleman:topological} that the representation in \eqref{eq:topological_charge} for $N(\n)$ was first derived in \cite{kleman:defect}. 

$N(\n)$ is additive: if the surface $\surface$ encloses more than one defect, the topological charge computed on it through \eqref{eq:topological_charge} is the algebraic sum of the topological charges computed on surfaces enclosing the single defects comprised in $\surface$. As pointed out in Sect.~VII.E.3 of \cite{mermin:topological}, a defect with topological charge $N(\n)$ can be transformed \emph{continuously} into a defect with \emph{opposite} topological charge, thus making $|N(\n)|$, and not $N(\n)$ itself, a topological invariant apt to classify point defects for director fields on $\sphere$. 

The mapping
\begin{equation}
	\label{eq:parity_transformation}
	\nR\mapsto\parity{\nR}:=-\nH
\end{equation} 
was described in \cite{sonnet:reorientational} as a \emph{parity} transformation, as it changes the sign of the topological charge,\footnote{This equation follows from Appendix~\ref{sec:topological_charge} and the general property of \eqref{eq:topological_charge} stating that $N(-\n)=-N(\n)$, which stems from being $(\nablas\n)^*$ even in $\n$.}
\begin{equation}
	\label{eq:parity_charge}
	N(\parity{\nR})=-N(\nR)=-1.
\end{equation}
This was meant to identify $\parity{\nR}$ as an \emph{anti} radial hedgehog, which would neutralize the topological charge of the radial hedgehog and annihilate it when combined together in a director field on $\sphere$ with zero total topological charge.\footnote{Actually, in \cite{sonnet:reorientational}, $\nH$ was defined to be precisely $\parity{\nR}$, so that, being opposite to the field in \eqref{eq:hip_hedg}, would form with it a defect-anti-defect pair, as would also be clear from Fig.~\ref{fig:radial_hyperbolic_hedgehogs} once the field lines  orientation in panel (b)  are reversed.} 

Here, instead, as shown in Appendix~\ref{sec:topological_charge},
\begin{equation}
	\label{eq:parity_charge_instead}
	N(\nH)=N(\nR)=+1,
\end{equation}
so that $\nR$ and $\nH$ not only belong to the same topological class, but also have the same topological charge.

We consider as domain $\body$ a ball $\ballx$ with radius $R$ and center at $\xv_0$. We study a trial \emph{twisted} hedgehog field $\nT$, which ``interpolates'' in space between $\nR$ and $\nH$. Formally, $\nT$ is obtained by acting on the radial hedgehog $\nR(\xv)$ in \eqref{eq:rad_hedg} with a rotation $\R(\alpha)$ of variable angle $\alpha=\alpha(r)$ about a fixed axis $\e\in\sphere$, where $r$ is the distance of $\xv$ from the defect at $\xv_0$,
\begin{subequations}\label{eq:twisted_hedgehog}
\begin{align}
	\nT(\xv)&:=\bm R(\alpha(r))\nR(\xv),\label{eq:twisted_hedgehog_formula}\\ \R(\alpha)&:=\I+\sin\alpha\W(\e)+(1-\cos\alpha)\W(\e)^2.\label{eq:rotation_alpha}
\end{align}
\end{subequations}
In \eqref{eq:rotation_alpha}, $\W(\e)$ is the skew-symmetric tensor associated with $\e$, whose action on any vector $\vv$ is given by $\W(\e)\vv=\e\times\vv$. The field $\nT$ reduces to the radial hedgehog $\nR$ for $\alpha\equiv 0$ and to the hyperbolic hedgehog $\nH$ for $\alpha\equiv\pi$.  As shown in Appendix~\ref{sec:topological_charge}, the topological charge of $\nT$ equals that of both $\nR$ and $\nH$, irrespective of the function $\alpha$,
\begin{equation}
	\label{eq:topological_charge_n_T}
	N(\nT)=+1.
\end{equation}
We shall call $\alpha$ the \emph{twist} angle.

\subsection{Inversion Ring}\label{sec:ring}
\begin{figure}[h]
	\centering 
	\includegraphics[width=.5\linewidth]{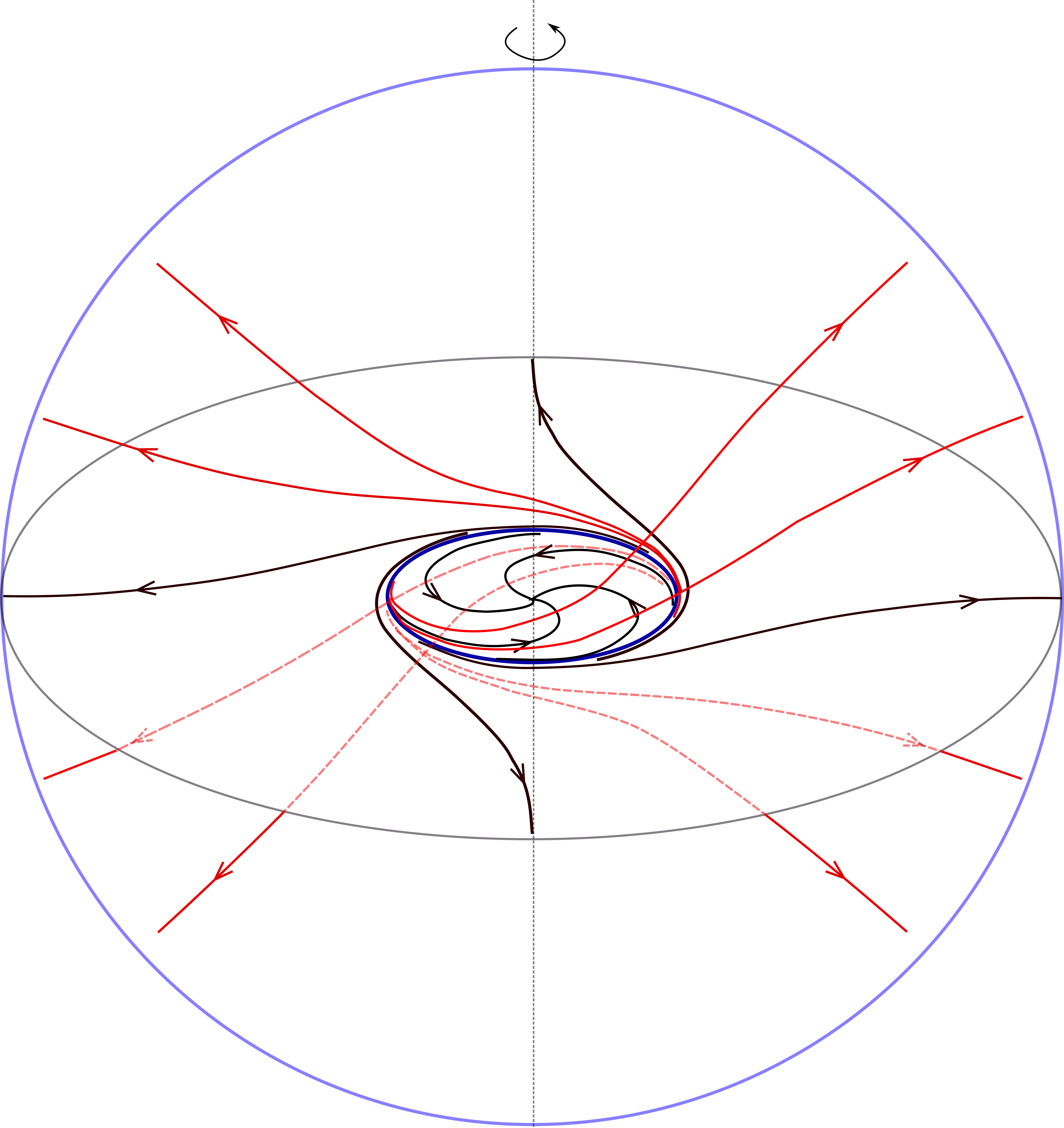}
	\caption{Field lines of $\nT$ in \eqref{eq:twisted_hedgehog} within the ball $\ballx$ enforcing condition \eqref{eq:condition_alpha_0}, so that $\nT=\nR$ on $\partial\ballx$. An inversion ring is present, which is depicted in blue. Black lines are field lines lying on the equatorial plane; red lines are field lines coming out of the equatorial plane. As in Fig.~\ref{fig:radial_hyperbolic_hedgehogs}, the whole 3D picture is obtained ny rotating this drawing about the polar axis.}
	\label{fig:twisted_hedgehog}
\end{figure}

A peculiar property of the field $\nT$ is illustrated by letting $\e$ be the polar axis of a  system of spherical coordinates $(r,\vt,\vp)$ with origin at  $\xv_0$. On the equatorial plane $\vt=\frac{\pi}{2}$, in the coordinate frame $(\e_r,\et,\ep)$, $\nT$ reduces to (see \eqref{eq:n_T_spherical_representation})
\begin{equation}\label{eq:n_T_equatorial_plane}
\nT=\cos\alpha\e_r+\sin\alpha\ep,	
\end{equation}	
and so it lies entirely on the equatorial plane. If, for some $r^*$, $\alpha(r^*)=\frac{\pi}{2}$, then $\nT$ is tangent to the circle of radius $r^*$ around $\xv_0$. Where $\alpha(r)<\frac{\pi}{2}$ the field $\nT$ spirals outward (relative to $\xv_0$), where $\alpha(r)>\frac{\pi}{2}$ it spiral inward. Figure~\ref{fig:twisted_hedgehog} illustrates this feature within the ball $\ballx$  when the condition
\begin{equation}
	\label{eq:condition_alpha_0}
	\alpha(R)=0
\end{equation}
is enforced, so that $\nT=\nR$ on the boundary $\partial\ballx$.
The ring at $r=r^*$ separates two opposite spiraling regimes; there, the field lines of $\nT$ appear to coalesce in a ring, which looks like a disclination, but is instead \emph{regular}, as it bears no discontinuity of the director. We shall call the ring at $r=r^*$, if present, an \emph{inversion} ring, as it marks the inversion of the spiraling sense.

It is perhaps the seminal work of Lavrentovich and Terentjev~\cite{lavrentovich:phase} where a first experimental evidence of an inversion ring within a twisted hedgehog was ever found and documented in ordinary nematics.\footnote{In a temperature regime where the twist constant $K_{22}$ is sufficiently small.}

Here, we shall use $\nT$ as a trial field to describe the twisted distortion that replaces $\nR$ in $\ballx$  when $\nR$ becomes unstable. We shall determine the function $\alpha$ subject to \eqref{eq:condition_alpha_0} that minimizes the elastic free energy $\free$ in \eqref{eq:free_energy_functional} with $\body=\ballx$. We shall do so for either $W=\WOF$ in \eqref{eq:Frank_equivalent} and $W=\WQT$ in \eqref{eq:quartic_free_energy_density} to see whether the quadratic and quartic elastic theories for chromonics recalled in Sect.~\ref{sec:energetics} can be distinguished on the basis of the predictions they make about the occurrence of a twisted hedgehog and its inversion ring.

We start by considering under what conditions $\nR$ is locally stable for either theory. 

\subsection{Local Stability of Radial Hedgehog}\label{sec:local_stibility}
First, we observe that $\nR$ is a \emph{universal} solution, as it solves 
the equilibrium equation for all possible elastic free-energy functionals $\free$ in \eqref{eq:free_energy_functional} associated with a frame-indifferent density $W=W(\n,\nabla\n)$, see \cite{ericksen:general}. Thus, $\nR$ is an equilibrium configuration for $\free$, irrespective of $W$. Moreover, it was proved in \cite{cohen:weak} and \cite{kinderlehrer:second} that, when $W=\WOF$, $\nR$ is a local minimizer of $\free$ in the admissible class of director fields $\n$ with finite energy in $\ballx$ and such that 
\begin{equation}
	\label{eq:boundary_condition_n}
	\n|_{\partial\ballx}=\nR,
\end{equation}
provided that the following inequality is satisfied,\footnote{A result which was independently rediscovered in \cite{rudinger:twist}.}
\begin{equation}
	\label{eq:loc_stab_hedgehog}
	0<k_1<1+\frac{k_{3}}{8}.
\end{equation}
Here and below, the elastic constants will be scaled to $K_{22}$,
\begin{equation}
	\label{eq:scaled_constants}
	k_1:=\frac{K_{11}}{K_{22}},\quad k_3:=\frac{K_{33}}{K_{22}},\quad\text{and}\quad k_{24}:=\frac{K_{24}}{K_{22}}\quad\text{with}\quad K_{22}>0.
\end{equation}
This local stability result is based on the study of the second variation of $\free$ at $\n=\nR$; the latter is the \emph{same} for both $\WOF$ and $\WQT$, as these only differ by a quartic term that does \emph{not} affect the second variation of $\free$ at $\nR$, see Appendix~\ref{sec:second_variation}. 

It is remarked in \cite{kinderlehrer:recent} that when \eqref{eq:loc_stab_hedgehog} is violated the free-energy functional $\free$ with $W=\WOF$ subject to \eqref{eq:boundary_condition_n} admits  a \emph{continuum} of minimizers, all sharing the same energy. Since the proof of this result is based on frame-indifference only, it also  holds within our quartic twist theory where $W=\WQT$. 

\begin{figure}[h]
	\centering 
	\includegraphics[width=.6\linewidth]{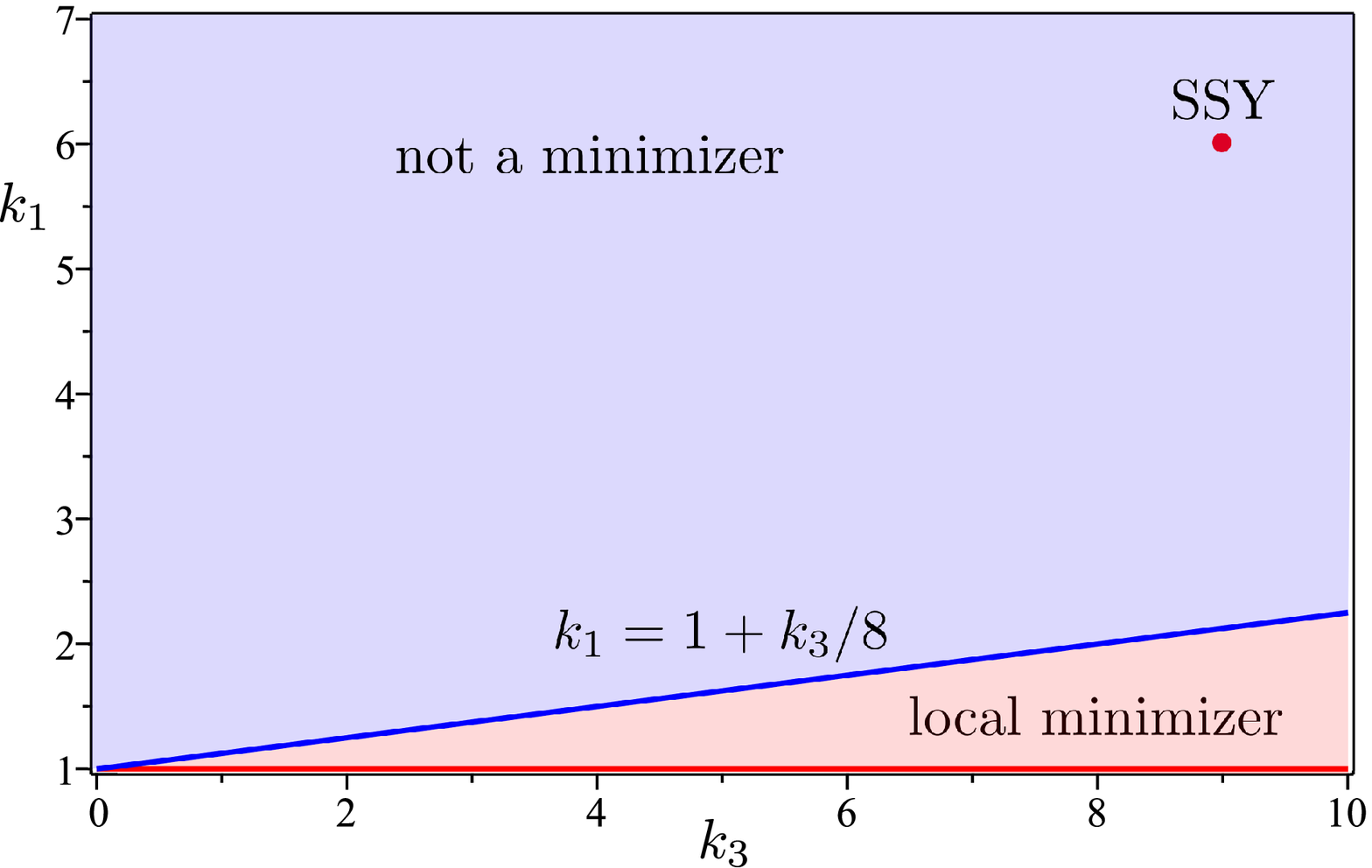}
	\caption{Regions of interest for the local stability of the radial hedgehog $\nR$ in CLCs. In the pink region, $\nR$ is a local minimizer of $\free$ subject to \eqref{eq:boundary_condition_n} when $\body=\ballx$, for either $W=\WOF$ and $W=\WQT$. In the blue region, $\nR$ is no longer a minimizer; there is a continuum of minimizers, all with the same energy. Bulk elastic constants of chromonics fall in the region of instability; the red dot represents data for SSY, $k_1\approx6.1$ and $k_3\approx8.7$, taken from \cite{zhou:elasticity_2012}.}
	\label{fig:graph_sum}
\end{figure}

Figure~\ref{fig:graph_sum} illustrates  inequality \eqref{eq:loc_stab_hedgehog} for $k_1>1$, which is the situation that applies to chromonics, as also shown by the dot representing data for SSY.
Hereafter, we assume that
\begin{equation}
	\label{eq:instability_hedg}
	k_1>1+\frac{k_3}{8}.
\end{equation}
The special family of twisted director configurations described by $\nT$ are parameterized by the scalar function $\alpha$ and the symmetry axis  $\e\in\mathbb{S}^2$. Once, for a given $\e$, $\alpha$ is chosen so as to minimize $\free$, letting $\e$ vary in $\sphere$  potentially embodies the  continuum of minimizers expected to arise when the radial hedgehog $\nR$ is no longer locally stable. 

\subsection{Minimum Problem}\label{sec:minimum_problem}
Here, we study the problem of minimizing the functional $\free$ in \eqref{eq:free_energy_functional} for $\body=\ballx$ and $W=\WQT$ subject to \eqref{eq:boundary_condition_n}. We introduce the change of variables
\begin{equation}
	\label{eq:rho_def}
	r\mapsto\rho:=\frac{r}{R},
\end{equation}
which maps $[0,R]$ onto $[0,1]$. In the new variable, \eqref{eq:condition_alpha_0} becomes\footnote{We shall continue to adopt the same old  symbol for the function $\alpha$, even if it is expressed in the new variable.}
\begin{equation}
	\label{eq:hom_cond}
	\alpha(1)=0.
\end{equation}
Standard computations (deferred to Appendix \ref{sec:standard_comp}) show that for $\body=\ballx$ and $W=\WQT$ the functional $\free$ in \eqref{eq:free_energy_functional} can be given the following scaled form
\begin{align}
	\label{eq:free_energy_twisted}
	&\mathcal{F}_\lambda[\alpha]:=\frac{15\free[\nT]}{8\pi K_{22}R}-\mathcal{F}_\mathrm{R}\nonumber\\
	&\quad=\int_0^1\bigg\{g(\alpha(\rho))(\rho\alpha'(\rho))^2+2(k_1-1)f_0(\alpha(\rho))+\frac{32}{7}\frac{\lambda^2}{\rho^2}\bigg[\sum_{n=1}^4 f_n(\alpha(\rho))(-\rho\alpha'(\rho))^n\bigg]\bigg\}\dd \rho,
\end{align}
where a prime $'$ denotes differentiation, $\mathcal{F}_\mathrm{R}:=15(k_1-k_{24})$ is the scaled energy of the radial hedgehog, so that 
\begin{equation}
	\label{eq:free_energy_radial_hedg}
	\mathcal{F}_\lambda[0]=0,
\end{equation}
and  the functions $g$, and $f_n$ are defined as
\begin{subequations}\label{eq:gf_free_energy}
	\begin{align}
		g(\alpha)&=2k_1\sin^2\alpha+\frac{2}{7}(1-\cos\alpha)^2+\frac{k_3}{14}\left(24\cos^2\alpha+8\cos\alpha+3\right),\label{eq:g_func}\\
		f_0(\alpha)&=2\cos^2\alpha+\cos\alpha-3,\label{eq:f_func}\\
		f_1(\alpha)&=3(1-8\cos \alpha)\sin^3\alpha,\label{eq:f_1}\\
		f_2(\alpha)&=(1-\cos \alpha)^2\sin^2\alpha,\label{eq:f_2}\\
		f_3(\alpha)&=\frac{2}{11}(1-\cos \alpha)^3\sin \alpha,\label{eq:f_3}\\
		f_4(\alpha)&=\frac{2}{143}(1-\cos \alpha)^4.\label{eq:f_4}
	\end{align}	
\end{subequations} 

 $\mathcal{F}_\lambda[\alpha]$ is invariant under the change of  $\alpha$ into $-\alpha$, for any $\alpha$. Thus, every  non-trivial equilibrium solution $\alpha_\lambda$ would be accompained by its parity conjugate $-\alpha_\lambda$. The corresponding fields $\nT$ differ as they have opposite chirality, but they have one and the same energy.

For $\lambda>0$, integrability of the quartic term in $\mathcal{F}_\lambda$ in \eqref{eq:free_energy_twisted} requires that the limiting value $\alpha(0)$ of $\alpha$ at $\rho=0$ be either $0$ or $\pi$. Under the assumption that, to within parity conjugacy, $\mathcal{F}_\lambda$ has a unique minimizer subject to \eqref{eq:hom_cond}, the choice $\alpha(0)=0$ would lead us to $\alpha\equiv0$, that is, to $\nR$, which is a contradiction since the radial hedgehog is unstable when \eqref{eq:loc_stab_hedgehog} applies.  Thus, we shall enforce the condition
\begin{equation}
	\label{eq:alpha_0}
	\alpha(0)=\pi\quad\text{for}\quad\lambda>0.
\end{equation}
For $\lambda=0$, $\alpha(0)$ is instead free to vary, as in \eqref{eq:free_energy_twisted} integrability is guaranteed by the integrability of $\alpha'$.\footnote{Which requires that $\rho\alpha'(\rho)$ be bounded as $\rho\to0^+$.}  

\subsubsection{Equilibrium Solutions}\label{sec:equilibrium_solutions}
Here, we specialize the analysis to  \emph{positive} solutions of the equilibrium equation for $\mathcal{F}_\lambda$: we assume that $\alpha_\lambda\geqq0$ since the minimizer of $\mathcal{F}_\lambda$ is not expected to change sign. Clearly, this positive branch of solutions remains associated with the conjugate negative branch, which has equal energy. The equilibrium equation is too complicated to lend itself to analytic solutions; it was symbolically manipulated and will be conventionally called (E).\footnote{It is equivalent to the equation of motion \eqref{eq:equation_of_motion_clcs} for the effective dynamical system described in Appendix~\ref{sec:eq_dyn_syst}.}

We could establish the asymptotic behaviour of the solutions $\alpha_\lambda$ of (E) near $\rho=0$ and $\rho=1$, for every $\lambda>0$. As shown  in Appendix~\ref{sec:asympt_rho0}, for $k_1>1$
\begin{equation}
	\label{eq:sol_asympt}
	\alpha_\lambda(\rho)= \pi(1-B\rho)+\mathcal{O}\left(\rho^2\right)\quad\text{for}\quad\rho\to0^+,
\end{equation}
where
\begin{equation}
	\label{eq:B_val}
	B=\sqrt{\frac{7}{32}}\frac{1}{\lambda}\frac{\sqrt{58058}\sqrt{21k_1 + 19k_3 - 5}}{1421\pi}.
\end{equation}
Similarly, 
\begin{equation}
	\label{eq:alpha_to_1}
	\alpha_\lambda(\rho)\approx C\left(\frac{1}{\rho}-1\right)\quad\text{as}\quad\rho\to1,
\end{equation}
where $C$ is a positive constant to be determined. 

\subsubsection{Energy Minimizers}\label{sec:energy_minimizers}
Here we explore numerically the minimizers $\alpha_\lambda$ of $\mathcal{F}_\lambda$, focusing on the positive equilibrium branch (thus selecting one chirality for $\nT$). For $\lambda=0$, this problem is solved in \cite{ball:brief} by reinterpreting $\mathcal{F}_0$ as an infinite-horizon action functional associated with an equivalent autonomous dynamical system in two-dimensional phase space. For $\lambda>0$, a similar reinterpretation for $\mathcal{F}_\lambda$ is still viable, but the associated dynamical system is \emph{not} autonomous; it is studied numerically in Appendix~\ref{sec:eq_dyn_syst} and contrasted with the autonomous system associated with $\mathcal{F}_0$.

The major difference between these dynamical systems resides in their equilibrium points; in the language of the twist angle $\alpha$, this translates into two different asymptotic values at the center of the ball $\ballx$,
\begin{equation}
	\label{eq:alpha_0_dichotomy}
	\alpha_\lambda(0)=
	\begin{cases}
	\widehat{\alpha}_0:=\arccos(-1/4)\quad&\text{for}\quad\lambda=0,\\
	\pi\quad&\text{for}\quad\lambda>0.	
	\end{cases}	
\end{equation}

One may say that the classical quadratic theory ($\lambda=0$) predicts that $\nR$ and $\nH$ are \emph{not} completely bridged inside the confining ball $\ballx$, whereas the quartic theory ($\lambda>0$) predicts that they are. Hence we could possibly use hedgehogs in chromonics confined within a ball to discriminate these theories from one another. However, although this is  a qualitative difference, its observation  might be experimentally precluded. A further, quantitative feature must be called upon; this is the size (relative to ball's radius $R$) of the inversion ring $r^*$ associated with the stable twisted hedgehogs predicted by both theories, as by \eqref{eq:alpha_0_dichotomy} an inversion ring is present in both cases.

\begin{figure}[h]
	\centering 
	\includegraphics[width=.4\linewidth]{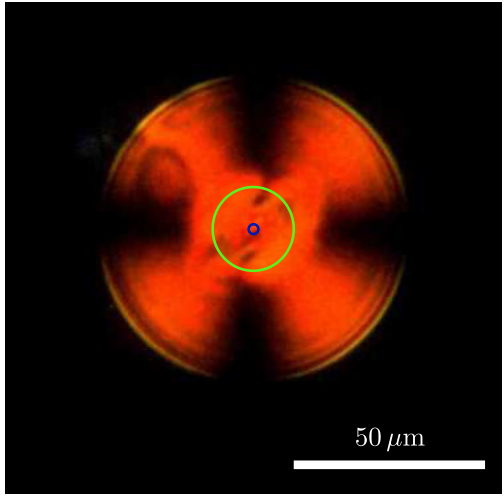}
	\caption{Reproduction of Fig.~5b of \cite{spina:intercalation} showing a spherical cavity (in a polymeric matrix) of radius $R\approx40.4\mu\mathrm{m}$ enclosing a SSY solution in water with concentration $30\%$ (wy/wt) and temperature $25\,\degree\mathrm{C}$. The homeotropic anchoring on the boundary of the sphere induces a (presumably twisted) hedgehog at the center exhibiting the typical Maltese cross when observed between crossed polarizers. The larger (green) and smaller (blue) circles superimposed to the figure are the inversion rings predicted by the quartic and quadratic theories, respectively. In absolute terms, with $a\approx6.4\mu\mathrm{m}$ (from \cite{paparini:elastic}), that is, $\lambda\approx0.16$, we have $r_0^*=\rho^*_0R\approx1\mu\mathrm{m}$ and $r^*_\lambda=\rho^*_\lambda R\approx8.1\mu\mathrm{m}$.}
	\label{fig:spina_figure}
\end{figure}

For definiteness, we consider a specific case, which was suggested by the experimental study in \cite{spina:intercalation}. This is the case of chromonic liquid crystal SSY in an aqueous solution (at a wt/wt concentration of $30\%$ and a temperature of $25\,\degree\mathrm{C}$) confined within a spherical cavity produced inside a polymeric matrix enforcing homeotropic anchoring for the director on its boundary (see Fig.~\ref{fig:spina_figure}). Material constants are derived from \cite{zhou:elasticity_2012} and deliver $k_1\approx6.1$ and $k_3\approx8.7$,\footnote{The absolute measured values are $K_{11}\approx4.3\,\mathrm{pN}$, $K_{22}\approx0.7\,\mathrm{pN}$, and $K_{33}\approx6.1\,\mathrm{pN}$.}   which, as shown in Fig.~\ref{fig:graph_sum}, locate the radial hedgehog in its unstable domain. The radius of the spherical cavity in Fig.~\ref{fig:spina_figure} is $R\approx40.4\mu\mathrm{m}$. For the same SSY solution in the same physical conditions,  in \cite{paparini:elastic} we estimated $a\approx6.4\mu\mathrm{m}$, thus here we take $\lambda=0.16$. 

\begin{figure}[h]
	\centering 
	\includegraphics[width=.6\linewidth]{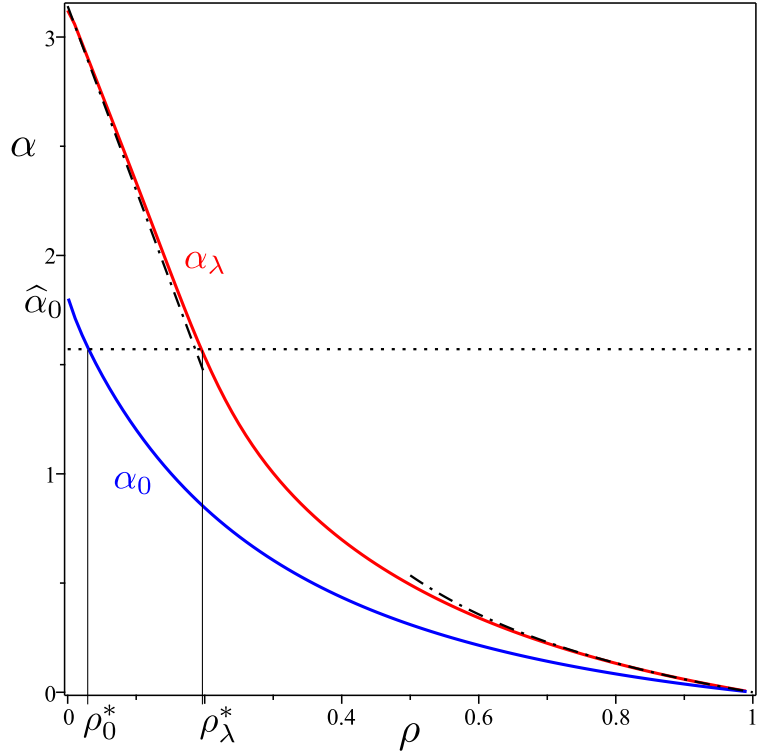}
	\caption{Plots against $\rho$ of the minimizer $\alpha_\lambda$ of $\mathcal{F}_\lambda$ (red curve) and the minimizer $\alpha_0$ of $\mathcal{F}_0$, for $k_1=6.1$, $k_3=8.7$, and $\lambda=0.16$. As in \eqref{eq:alpha_0_dichotomy}, $\widehat{\alpha}_0:=\arccos(-1/4)$. The broken lines reproduce the asymptotic behaviours predicted by \eqref{eq:sol_asympt} and \eqref{eq:alpha_to_1}, respectively, with $B\doteq2.68$, in agreement with \eqref{eq:B_val}, and $C\doteq0.54$. The dotted line drawn at $\alpha=\frac{\pi}{2}$ intercepts the graphs of $\alpha_\lambda$ and $\alpha_0$ at values $\rho^*$ of $\rho$ that designate the scaled radius $r^*$ of the inversion ring in the two cases. It is apparent how  $\rho^*_\lambda\approx0.2$ is appreciably larger than $\rho^*_0\approx0.03$, see also Fig.~\ref{fig:spina_figure} below.}
	\label{fig:alpha_profiles}
\end{figure}

The profile of the minimizing twist angle $\alpha_\lambda$ corresponding to these parameters is shown in Fig.~\ref{fig:alpha_profiles} (red curve) against the minimizing profile $\alpha_0$ for $\lambda=0$ (blue curve).
It is apparent that the inversion rings associated with these solutions are appreciably different.

Figure~\ref{fig:spina_figure} reproduces a spherical cavity (in a polymeric matrix) observed in \cite{spina:intercalation}; 
there we also draw the inversion rings predicted by both  classical and quadratic theories. Judging from this single comparison and taking for granted that the defect shown here is indeed a twisted hedgehog, we may say that the quartic theory seems to capture better the size of the inner  structure enclosed by the inversion ring.
This core structure  will be further detailed in Sect.~\ref{sec:cores}.

\begin{figure}[h]
	\centering 
	\includegraphics[width=.6\linewidth]{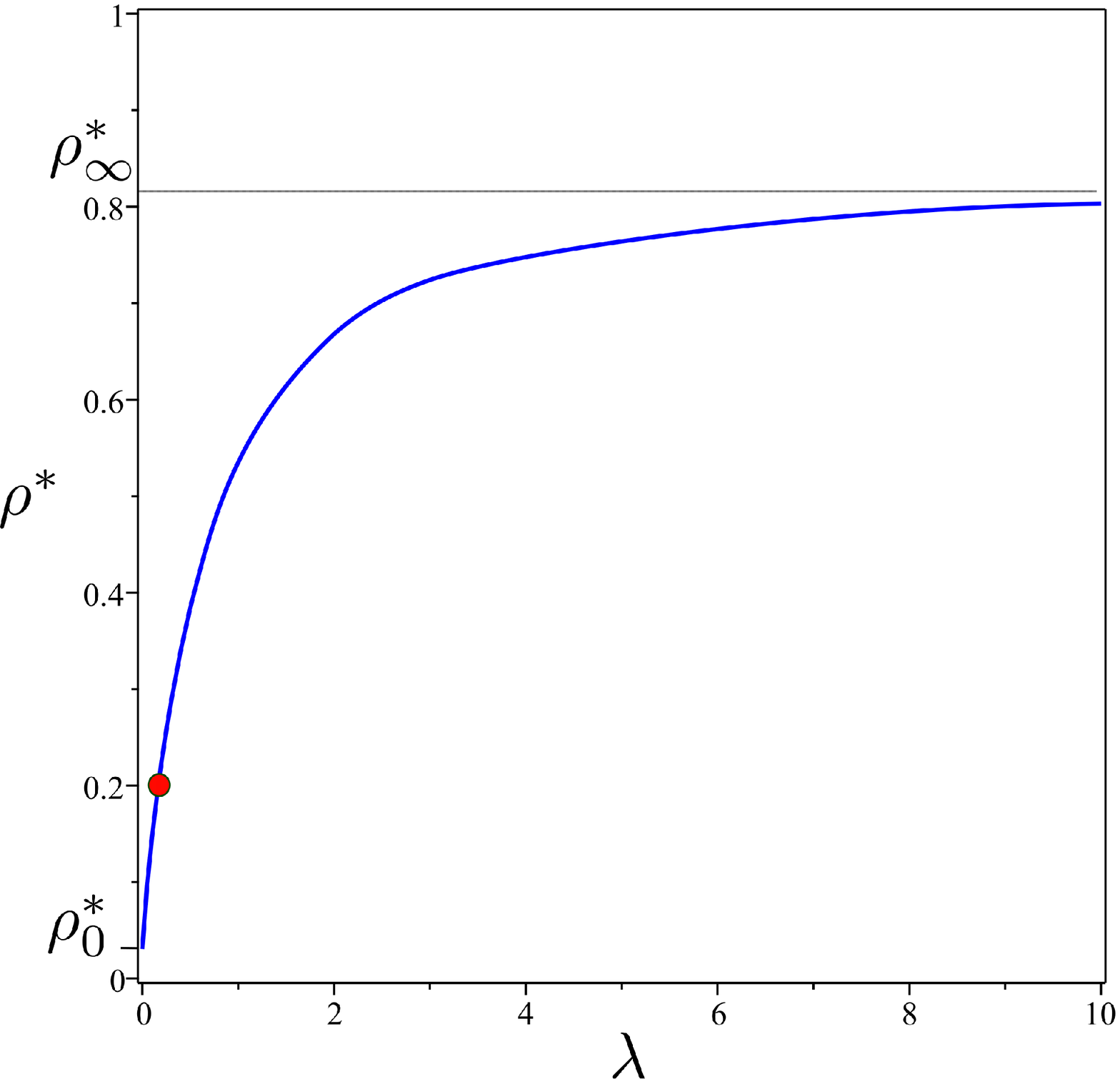}
	\caption{Plot of $\rho^*=r^*/R$ as a function of $\lambda$ computed on the minimizer $\alpha_\lambda$ of $\mathcal{F}_\lambda$; the graph saturates at $\rho^*_\infty\approx0.82$, while $\rho^*_0\approx0.03$ is the limiting value as $\lambda\to0$. The red dot marks the inversion ring predicted for $\lambda=0.16$, corresponding to the spherical cavity shown in Fig.~\ref{fig:spina_figure}.}
	\label{fig:rho_star_lambda}
\end{figure}

Letting $\rho^*:=r^*/R$ designate the scaled radius of the inversion ring, we explored the dependence of $\rho^*$ on $\lambda$. The plot in Fig.~\ref{fig:rho_star_lambda} summarizes the outcomes of this analysis; it shows how $\rho^*$ saturates to $\rho^*_\infty\approx0.82$ as $\lambda$ grows indefinitely.
  
Not only does the inversion ring size increase monotonically with $\lambda$, but also the defect core inside the inversion ring is qualitatively different for $\lambda=0$ and $\lambda>0$. These differences will be highlighted in the following section. 

\section{Spiraling Cores}\label{sec:cores}
Here we go into deeper details of the twisted hedgehog $\nT$ that minimizes the elastic free energy $\mathcal{F}_\lambda$; we are especially interested in the behaviour if its field lines within the defect core, which is conveniently identified with a sphere of radius $r^*$, the radius of the inversion ring. We shall again study primarily the distortion afforded by the quartic theory with $\lambda>0$; this case will also be contrasted against the case $\lambda=0$ of the classical quadratic theory. We shall see that the differences between the two cases are both qualitative and quantitative.

We split our analysis in two steps; in the first, we study the field lines of $\nT$ on the equatorial plane of $\ballx$ (orthogonal to the symmetry axis); in the second, we see how these lines behave away from that plane.

\subsection{Equatorial Field Lines}
In a spherical coordinate system $(r,\vt,\vp)$ with polar angle $\vt\in[0,\pi]$, the equatorial plane is described by $\vt=\frac{\pi}{2}$ and $r\geqq0$, $\vp\in[0,2\pi)$. Scaling lengths to the radius $R$ of the spherical cavity and letting $\rho$ be still defined as in \eqref{eq:rho_def} above, we see from \eqref{eq:n_T_equatorial_plane} that the field lines of $\nT$ on the equatorial plane are the solutions $(\vp(\tau),\rho(\tau))$ to the differential system
\begin{subequations}\label{eq:field_lines_equator_system+initial}
\begin{align}
		\frac{\dd\vp}{\dd\tau}&=1,\label{eq:field_lines_equator_system_phi}\\
		\frac{\dd\rho}{\dd\tau}&=\frac{\rho(\tau)}{\tan(\alpha_\lambda(\rho(\tau)))}\label{eq:field_lines_equator_system_rho},
\end{align}
subject to
\begin{equation}
	\label{eq:initial_conditions_equator_initial}
	\vp(0)=0\quad\text{and}\quad\rho(0)=\rho_0\quad\text{with}\quad0<\rho_0<1,
\end{equation}
\end{subequations}
where $\tau$ is a parameter.
The curves solving \eqref{eq:field_lines_equator_system+initial} may be winding several times around the origin as $\tau\to+\infty$; the appropriate solution of \eqref{eq:field_lines_equator_system_phi} is then
\begin{equation}
	\label{eq:phi_mod_2pi}
	\vp=\tau\quad\mathrm{mod}\ 2\pi.
\end{equation}

\begin{figure}
	\centering
	\subfloat[Quadratic theory (with $\lambda=0$): The inversion ring has (scaled) radius $\rho^*\approx0.03$. Zooming inside the inversion ring reveals the logarithmic nature of the asymptotic spirals.]{%
		\resizebox*{5.0cm}{!}{\includegraphics{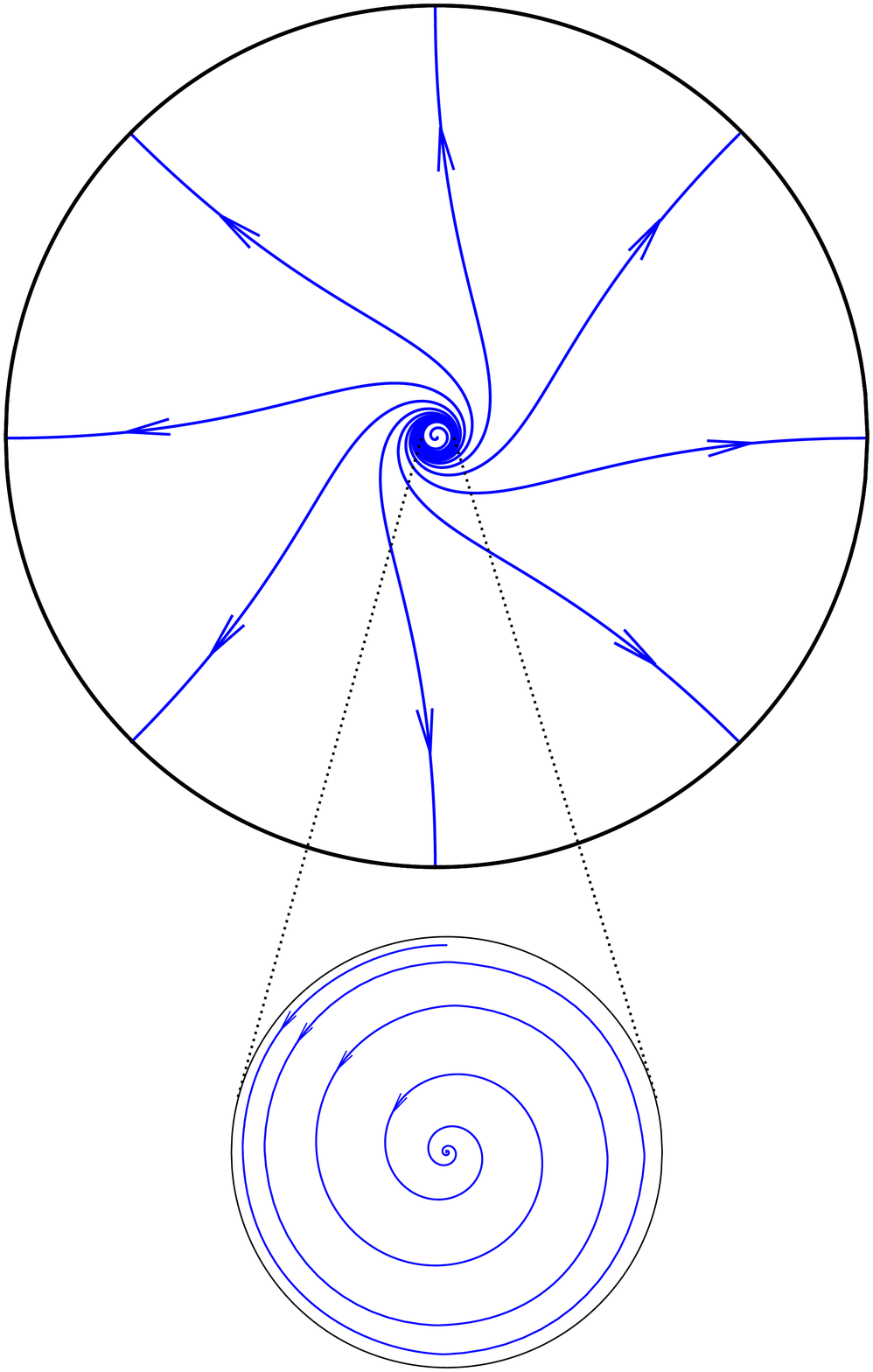}}}\hspace{25pt}
	\subfloat[Quartic theory (with $\lambda=0.16$): The inversion ring has (scaled) radius $\rho^*\approx0.2$. Zooming inside the inversion ring reveals the Archimedean nature of the asymptotic spirals.]{%
		\resizebox*{5.0cm}{!}{\includegraphics{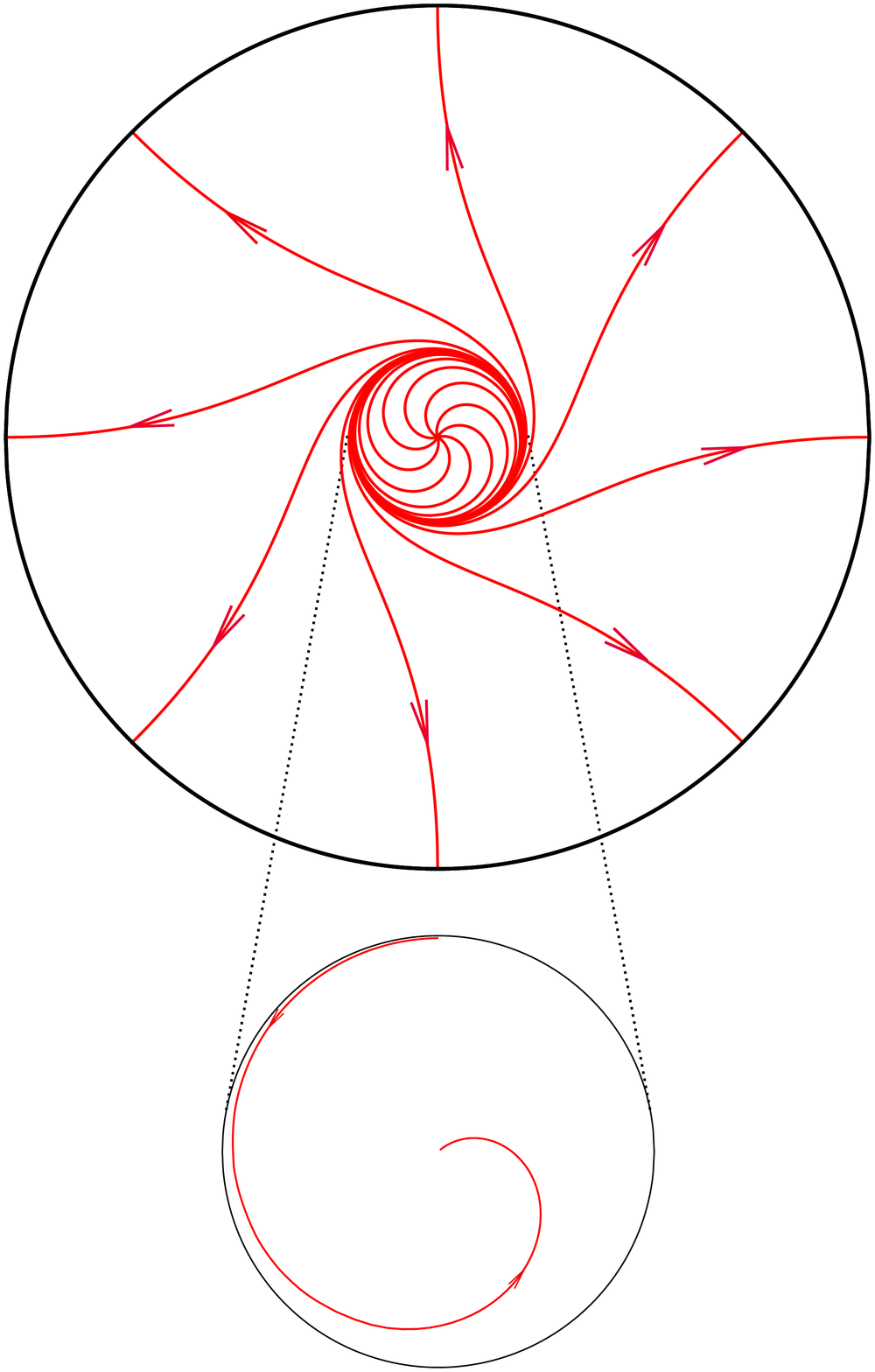}}}
	\caption{Field lines of $\nT$ in the equatorial plane of $\ballx$ according to the two elastic theories considered here. Material constants correspond to SSY in the same conditions that apply to both Figs.~\ref{fig:alpha_profiles} and \ref{fig:spina_figure}.}
	\label{fig:equatorial_integral_lines}
\end{figure}

It follows from \eqref{eq:field_lines_equator_system+initial} that every field line that starts inside or outside the inversion ring, remains inside or outside that ring, respectively. The inversion ring at $\rho=\rho^*$ is a field line  itself, since $\rho\equiv\rho^*$ is a solution of \eqref{eq:field_lines_equator_system_rho}. Moreover, a field line that starts   from $\rho_0<\rho^*$ keeps spiraling (clockwise) around the point defect at the origin,  while a field line that starts from $\rho_0>\rho^*$ is soon bent (anti-clockwise) towards the equator of $\ballx$, where it points radially away from the defect (see Fig.~\ref{fig:equatorial_integral_lines}).

Another qualitative feature of the field lines of $\nT$ is revealed by \eqref{eq:field_lines_equator_system+initial}.
Given the monotonicity of $\rho(\tau)$ both inside and outside the inversion ring, this function is invertible; a straightforward integration yields the following formula for its inverse,
\begin{equation}
	\label{eq:integral_lines_inversion}
	\tau(\rho)=
	\begin{cases}
\int_{\rho_0}^\rho \frac{\tan\alpha_\lambda(\xi)}{\xi}\dd\xi\quad&\text{for}\quad\rho>\rho^*,\\
\int_{\rho}^{\rho_0} \frac{\tan\alpha_\lambda(\xi)}{\xi}\dd\xi\quad&\text{for}\quad\rho<\rho^*.
	\end{cases}
\end{equation}
Two noteworthy consequences follow from \eqref{eq:integral_lines_inversion}. First, from the divergence of both integrals as $\rho\to\rho^*$ (from above and from below, respectively), we see that the field lines of $\nT$ wind infinitely many times around the inversion ring, no matter which elastic theory is employed to describe a twisted hedgehog. Second, by taking the limit as $\rho\to0^+$ in the second integral, we see that this diverges or not, depending on the limiting value $\alpha_\lambda(0)$. Since the latter depends on being $\lambda=0$  or $\lambda>0$, the two theories being compared here afford different qualitative predictions. According to the quadratic theory, for which $\alpha_0=\arccos(-1/4)$, the second integral in \eqref{eq:integral_lines_inversion} diverges and the field lines of $\nT$ wind infinite many times around the point defect at the origin; asymptotically, they are \emph{logarithmic} spirals. On the contrary, according to the quartic theory, for which $\alpha_\lambda=\pi$ for all $\lambda>0$, by \eqref{eq:sol_asympt}, the second integral in \eqref{eq:integral_lines_inversion} converges and the field lines of $\nT$ wind a finite number of times around the defect; asymptotically, they are \emph{Archimedean} spirals.  

In Fig.~\ref{fig:equatorial_integral_lines}, the field lines of $\nT$ in the equatorial plane are contrasted for the two theories, when the twist angle is given by the functions $\alpha_0$ and $\alpha_\lambda$ whose graphs are shown in Fig.~\ref{fig:alpha_profiles}. In both cases the inversion ring is zoomed in to highlight the different nature of the asymptotic spirals around the point defects.

\subsection{Field Lines in Space}
As is easily seen from \eqref{eq:n_T_spherical_representation}, the field lines of $\nT$ away from the equatorial plane of $\ballx$ are described in spherical coordinates $(r,\vt,\vp)$ by the solutions to the following differential system
\begin{subequations}\label{eq:prime_rho_phi}
\begin{align}
		\frac{\dd\rho}{\dd\tau}&=\rho(\tau)\dfrac{1+(\cos\alpha_\lambda(\rho(\tau))-1)\sin^2\vt}{\sin\alpha_\lambda(\rho(\tau))},\\
		\frac{\dd\vt}{\dd\tau}&=\frac{(\cos\alpha_\lambda(\rho(\tau))-1)\cos\vt\sin\vt}{\sin\alpha_\lambda(\rho(\tau))},\\
		\frac{\dd\vp}{\dd\tau}&=1,
\end{align}
\end{subequations}
where $\tau$ is a parameter chosen again so that \eqref{eq:phi_mod_2pi} holds. 

\begin{figure}[h]
	\centering
	\includegraphics[width=.9\linewidth]{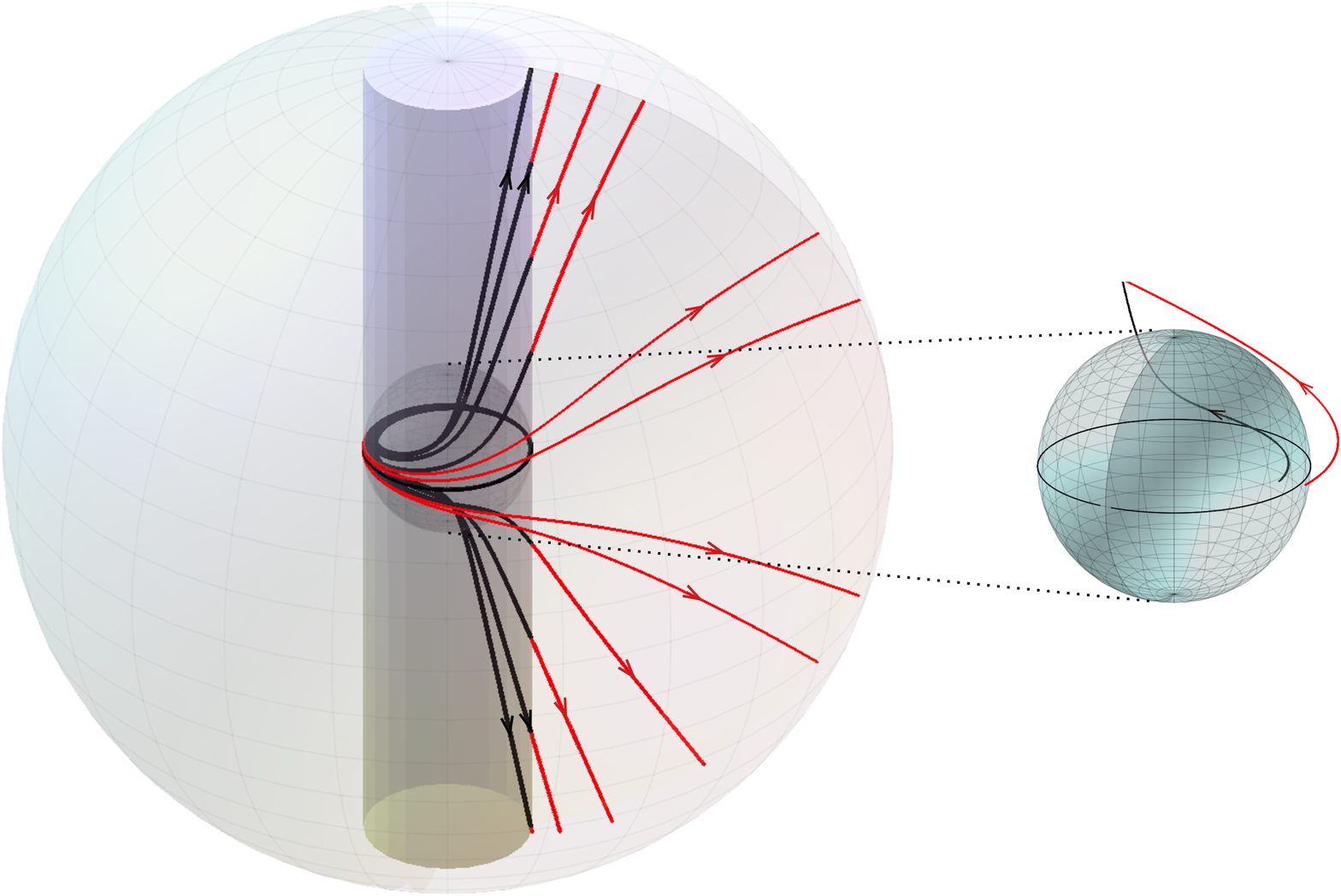}
	\caption{Field lines of $\nT$ away from the equatorial plane of $\ballx$, for the same choice of parameters in both Figs.~\ref{fig:alpha_profiles} and \ref{fig:spina_figure}. Only the two limiting negatively invariant sets, the ball $\ball_{r^*}$ and the cylinder $\cyl_{r^*}$ built on the inversion ring, are shown. Field lines are back inside $\cyl_{r^*}$ and red outside. The zoomed region on the right is the ball of radius $r^*$; two field lines are drawn that start near the boundary of $\ball_{r^*}$, one inside $\cyl_{r^*}$ (black) and the other outside (red).}
	\label{fig:space_integral_lines_3D}
\end{figure}

The flow described by \eqref{eq:prime_rho_phi} is mirror-symmetric with respect to the equatorial plane ($\vt=\frac{\pi}{2}$) and, as shown in Fig.~\ref{fig:space_integral_lines_3D},
possesses two families of \emph{negatively} invariant sets, balls and circular cylinders with radii larger than the radius $r^*$ of the inversion ring. This means that field lines of $\nT$ may only leave the regions enclosed by these sets and never enter them.

\begin{figure}[h!] 
	\centering
	\includegraphics[width=.75\linewidth]{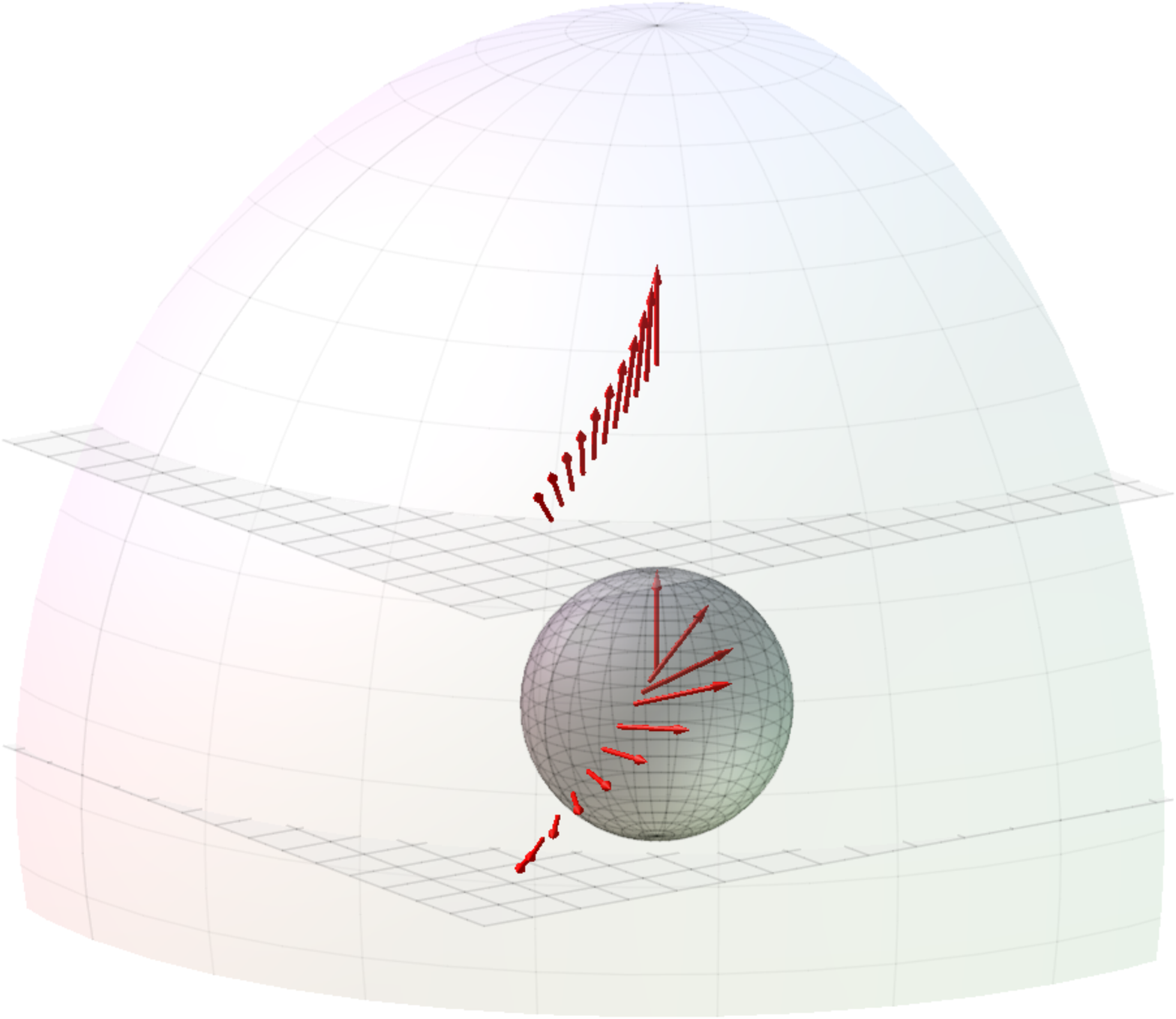}
	\caption{For the same field $\nT$ in Fig.~\ref{fig:space_integral_lines_3D}, the director profiles are shown on two parallel sections of $\ballx$ with planes parallel to the equator: one cuts the ball $\ball_{r^*}$ at mid-height, $z=r^*/2$, while the latter cuts the ball $\ballx$ at mid-height, $z=R/2$.}
	\label{fig:sec_z_const}
\end{figure}

To prove this qualitative property, we denote by $\ball_r$ and $\cyl_r$  these families of  balls and cylinders, respectively, and by $\normal$ their outer unit normal. It readily follows from \eqref{eq:n_T_spherical_representation} that
\begin{subequations}
\begin{align}
	\nT\cdot\normal|_{\partial\ball_r}&=\sin^2\vt\cos\alpha+\cos^2\vt,\\
		\nT\cdot\normal|_{\partial\cyl_r}&=\sin\vt\cos\alpha,
\end{align}
\end{subequations}
which are both non-negative for all $\vt\in[0,\pi]$ whenever $\alpha\leqq\frac{\pi}{2}$, that is, for $r>r^*$. A further geometric illustration of this  property is given in Fig.~\ref{fig:sec_z_const}.

\section{Conclusion}\label{sec:discussion}
In \cite{paparini:elastic}, we proposed a quartic twist theory for the curvature elasticity of chromonic liquid crystals, for which we have been seeking corroborating evidence. This theory introduces a phenomenological length $a$, which in \cite{paparini:elastic} was estimated to be of the order of microns by fitting published data for chromonics filling cylinders with degenerate planar anchoring on their lateral boundary. These data could also be interpreted by use of the classical quadratic Oseen-Frank theory \cite{davidson:chiral,nayani:spontaneous}, which however would be unable to predict stable shapes for the tactoidal droplets observed in the biphasic region  of these materials \cite{paparini:paradoxes}.

We turned to hedgehog defects and their core structure to find an instance where the two theories would afford different predictions, which could serve to differentiate them. We considered a spherical cavity of radius $R$ enforcing homeotropic anchoring on its boundary, like those produced in \cite{spina:intercalation}, and studied the \emph{twisted} hedgehogs predicted by both theories in the region in parameter space where the radial hedgehog would be unstable. 

The defect core of a twisted hedgehog director field $\nT$ is characterized by an inversion ring that encloses the defect core. Two properties of the defect core are predicted in stark contrast by the two theories: one is qualitative, the other quantitative.

We start with the latter. The radius $r^*$ of the inversion ring depends only on the elastic anisotropy for the quadratic theory and also on the ratio $\lambda=a/R$ for the quartic theory. For SSY in the same physical conditions as in \cite{spina:intercalation}, taking $a$ from \cite{paparini:elastic}, we estimated $r^*$ to be nearly an order of magnitude larger for the quartic theory compared to the quadratic one, $8.1\,\mu\mathrm{m}$ against $1\,\mu\mathrm{m}$.

On the qualitative side, we showed that the field lines of $\nT$ spiral differently around the point defect according to which theory is adopted: in  the quadratic theory, they are logarithmic spirals; in  the quartic theory, they are instead Archimedean spirals.

We may perhaps say that the defect core of  twisted defects, with its distinctive quantitative and qualitative features, could be the hallmark of a quartic elastic theory for chromonics. However, such a clear distinction between quadratic and quartic theories rests on being $a$ in the order of microns; were it much smaller, the differences highlighted here could not be appreciated. A thorough  study with direct observations of the core structure of twisted hedgehogs would be desirable.

Another critical issue that deserves further research concerns the splay constant $K_{11}$. If the recent theoretical  estimate for the  elastic constants in 
\cite{ravignas:spontaneous} is to be confirmed by different, independent approaches, not only $K_{22}$, but also $K_{11}$ would be smaller than $K_{24}$ for chromonics. This, as shown in \cite{paparini:paradoxes}, would ignite the instability of chromonic droplets in an isotropic  fluid environment enforcing  homeotropic anchoring at the interface. The defects studied in this paper inhabit a spherical cavity of \emph{fixed} shape, and so they are saved from that  instability. However, should homeotropic anchoring be realistic for chromonic droplets, if $K_{11}<K_{24}$, our quartic \emph{twist} theory could \emph{not} prevent such a shape instability, as it would be driven by a concentration of splay. Thus, were homeotropic chromonic droplets actually observed, our elastic theory would need to be amended. 

\appendix

\section{Trial Twisted Hedgehog}\label{sec:trial_twisted}
This Appendix contains ancillary results instrumental to our analysis in the main text.

\subsection{Useful Computations}\label{sec:standard_comp}
Identifying  the the unit vector $\e$ designating in \eqref{eq:rotation_alpha} the symmetry axis of $\nT$ as the polar axis $\e_z$ of standard spherical coordinates   $(r,\vt,\vp)$, where $\vt\in[0,\pi]$ is the polar angle and $\vp\in[0,2\pi)$ is the azimuthal angle,  we represent 
the gradient of the trial twisted field through the formula
\begin{align}\label{eq:nabla_twisted}
	\nabla\nT=&\frac{1}{r}\left[\Pr+\sin\alpha\Wz+(1-\cos\alpha)\Wz^2\right]+\left(\alpha'\cos\alpha-\frac{1}{r}\sin\alpha\right)\Wz\e_r\otimes\e_r\nonumber\\
	+&\left[\alpha'\sin\alpha-\frac{1}{r}(1-\cos\alpha)\right]\Wz^2\e_r\otimes\e_r,
\end{align}
where $\Pr:=\I-\e_r\otimes\e_r$ is the projection onto the plane orthogonal to $\e_r$, $\Wz$ is the skew-symmetric tensor with axial vector $\e_z$, and  a prime $'$ denotes differentiation with respect to $r$.

The following expressions for the traditional measures of distortion of $\nT$ in \eqref{eq:twisted_hedgehog_formula} are consequences of \eqref{eq:nabla_twisted}; they are written in the local frame $(\e_r,\e_\vt,\e_\vp)$ of spherical coordinates:
\begin{subequations}\label{eq:distortion_measures}
	\begin{align}
		\diver\nT&=\frac{1}{r}\left[-(r\alpha')\sin\alpha\sin^2\vt+1-(1-\cos\alpha)\cos^2\vt+\cos\alpha\right],\label{eq:div}\\
		\curl\nT&=\frac{1}{r}\left\{2\sin\alpha\cos\vt\e_r-\sin\vt\left[(r\alpha')\cos\alpha+\sin\alpha\right]\e_\vt\right.\nonumber\\
		&\left.+\cos\vt\sin\vt\left[-(r\alpha')\sin\alpha+(1-\cos\alpha)\right]\e_\vp\right\},\label{eq:curl}\\
		\nT\cdot\curl\nT&=\frac{1}{r}\left\{\cos\vt\left[-(r\alpha')(1-\cos\alpha)\sin^2\vt+2\sin\alpha\right]\right\},\label{eq:twist}\\
		\nT \times \curl\nT&=\frac{1}{r}\{\sin^2\vt[(r\alpha')\sin\alpha(1-(1-\cos\alpha)\sin^2\vt)\nonumber\\ &\qquad\qquad-(1-\cos\alpha)^2\cos^2\vt+\sin^2\alpha]\e_r\nonumber\\
		&\qquad+\sin\vt[-(r\alpha')\cos\alpha(1-(1-\cos\alpha)\sin^2\vt)\nonumber\\
		&\qquad\qquad\quad+\sin\alpha(-1+(1-\cos\alpha)(1+\cos^2\vt))]\et\}\nonumber\\
		&\qquad+\sin\vt\cos\vt[(r\alpha')\sin\alpha(1-(1-\cos\alpha)\sin^2\vt)\nonumber\\
		&\qquad\qquad\qquad-(1-\cos\alpha)(1-(1-\cos\alpha)\sin^2\vt)+2\sin^2\alpha]\e_\vp,\label{eq:bend}\\
		\tr(\nabla\nT)^2-(\diver\nT)^2&=-2(\cos^2\vt-(r\alpha')\sin\alpha\sin^2\vt+\cos\alpha\sin^2\vt).\label{eq:saddle_splay}
	\end{align}	
\end{subequations} 
Making use of \eqref{eq:distortion_measures} and \eqref{eq:hom_cond} in the free energy density $\WQT$ in \eqref{eq:quartic_free_energy_density}, and integrating over $\body=\ballx$, we arrive at the following scaled form for $\free$ in \eqref{eq:free_energy_functional},
\begin{equation}
	\label{eq:free_energy_twisted0}
\frac{15\free[\nT]}{8\pi K_{22}R}=:\mathcal{F}_\lambda[\alpha]+\mathcal{F}_\mathrm{R},
\end{equation}
where $\mathcal{F}_\lambda[\alpha]$ and $\mathcal{F}_\mathrm{R}$ are given by  \eqref{eq:free_energy_twisted} and 
\begin{equation}
	\label{eq:F_scaled_R}
	\mathcal{F}_\mathrm{R}=15(k_1-k_{24}),
\end{equation}
respectively.

\subsection{Topological Charge}\label{sec:topological_charge}
Here we compute the topological charge $N(\nT)$ of the twisted hedgehog $\nT$ in \eqref{eq:twisted_hedgehog}. To this end, we first note that 
\begin{equation}
	\label{eq:nabla_n_T_first_step}
\nablas\nT=(\nabla\n)\Pr=\frac{1}{r}[\Pr+\sin\alpha\Wz\Pr+(1-\cos\alpha)\Wz^2\Pr],
\end{equation}
where use has been made of \eqref{eq:nabla_twisted}. In the frame $(\e_r,\e_\vt,\e_\vp)$, $\nablas\nT$ in \eqref{eq:nabla_n_T_first_step} is also represented as
\begin{align}\label{eq:nabla_n_T_second_step}
\nablas\nT=\frac{1}{r}\{&(\sin^2\vt+\cos^2\vt\cos\alpha)\et\otimes\et-\cos\vt\sin\alpha\et\otimes\ep\nonumber\\
&+\cos\vt\sin\alpha\ep\otimes\et+\cos\alpha\ep\otimes\ep\nonumber\\
&-\sin\vt\cos\vt(1-\cos\alpha)\e_r\otimes\et-\sin\vt\sin\alpha\e_r\otimes\ep\},
\end{align}	
where we have employed the identity
\begin{equation}
	\label{eq:identity}
	\et=\frac{1}{\sin\vt}(\cos\vt\e_r-\e_z),
\end{equation}
having identified $\e$ and $\e_z$, as above.

It is now a simple matter to compute in the basis $(\e_r,\et,\ep)$ the tensor $(\nablas\nT)^*$, as it is represented by the cofactor matrix of the matrix representing $\nablas\nT$ in \eqref{eq:nabla_n_T_second_step}. A tedious, but simple calculation delivers 
\begin{equation}
	\label{eq:nablas_n_T_cofactor}
	(\nablas\nT)^*=\frac{1}{r^2}\{\sin\vt\cos\vt(\cos\alpha-1)\et\otimes\e_r+\sin\vt\sin\alpha\ep\otimes\e_r+(\sin^2\vt\cos\alpha+\cos^2\vt)\e_r\otimes\e_r\}.
\end{equation}
Since it follows from \eqref{eq:twisted_hedgehog} and \eqref{eq:identity} that
\begin{equation}
	\label{eq:n_T_spherical_representation}
	\nT=\cos\vt\sin\vt(\cos\alpha-1)\et+\sin\vt\sin\alpha\ep+(\sin^2\vt\cos\alpha+\cos^2\vt)\e_r,
\end{equation}
it is easily concluded that 
\begin{equation}
	\label{eq:n_T_pre_charge}
	\nT\cdot(\nablas\nT)^*\e_r=\frac{1}{r^2}.
\end{equation}
Taking $\surface$ in \eqref{eq:topological_charge}  to be a sphere of radius $r$ and center at $\xv_0$, we readily obtain that $N(\nT)=+1$, for any function $\alpha$, which is precisely \eqref{eq:topological_charge_n_T} in the main text. In particular, by taking $\alpha\equiv0$ or $\alpha\equiv\pi$, we recover \eqref{eq:parity_charge_instead}.

\subsection{Second Variation}\label{sec:second_variation}
We let $\free_4$ denote the quartic term contribution to $\free$ in \eqref{eq:free_energy_functional} arising from $\WQT$ in \eqref{eq:quartic_free_energy_density},
\begin{equation}
	\label{eq:quartic_free_energy_integral}
	\free_4[\n]:=\frac14K_{22}a^2\int_{\body}(\n\cdot\curl\n)^4\dd V.
\end{equation}
By applying to $\free_4$ the method illustrated in \cite{paparini:stability}, we readily see that the second variation $\delta^2\free_4$ of $\free_4$ at $\n$ can be given the general form 
\begin{align}
	\label{eq:second_variation}
	\delta^2\free_4(\n)[\vv]=K_{22}a^2\int_{\body}\Big\{&(\n\cdot\curl\n)^2\Big[3(\vv\cdot\curl\n+\n\cdot\curl\vv)^2\nonumber\\&+2(\n\cdot\curl\n)(v^2\n\cdot\curl\n+\vv\cdot\curl\vv)\Big]\Big\}\dd V,
\end{align}
which is a quadratic functional in the perturbation field $\vv$ subject to the orthogonality condition
\begin{equation}
	\label{eq:orthogonality_condition}
	\n\cdot\vv\equiv0.
\end{equation}
It is a very simple matter to check that $\delta^2\free_4(\nR)\equiv0$, as $\curl\nR\equiv\zero$.
\subsection{Asymptotic Behaviours}\label{sec:asympt_rho0}
Here we give a few details about the derivation of the asymptotic behaviours in \eqref{eq:sol_asympt} and \eqref{eq:alpha_to_1} of the equilibrium solutions $\alpha_\lambda$ for $\mathcal{F}_\lambda$. We renounce writing the equilibrium equation of $\mathcal{F}_\lambda$ since it is too complicated; as in the main text, it will denoted by (E) and manipulated by symbolic calculus. An equivalent form of (E) will be encountered in Appendix~\ref{sec:eq_dyn_syst} below.

When $\rho$ is near $0$, we write $\alpha_\lambda$ as
\begin{equation}
	\label{eq:beta_asympt_quartic}
	\alpha_\lambda(\rho)\approx\pi\left(1-B\rho^{\beta}\right),
\end{equation}
which satisfies \eqref{eq:alpha_0}, and seek $B>0$ and  $\beta>1/4$, assumed to exist, the latter requirement being a direct consequence of the integrability of $\mathcal{F}_\lambda$. In our asymptotic method, which is an adaptation of the classical method of Frobenious (see, for example, p.\,396 of \cite{ince:ordinary}), we determine both $\beta$ and $B$ by requiring that the dominant term of (E) vanishes. The first two powers of (E) near $\rho=0$ are as follows 
\begin{subequations}
\begin{equation}
	\label{eq:EL_bulk_hedg_asympt}
	B^2\pi^2\lambda^2P_4(\beta)\rho^{-2 + 3\beta}
	+\frac{143}{64}P_2(\beta)\rho^{\beta},
\end{equation}
where the polynomials $P_2$ and $P_4$ are defined as
\begin{align}\label{eq:polynomials}
	P_2(\beta)&:=\left(\frac{19k_3}{42} + \frac{8}{21}\right)\beta^2 + \left(\frac{19k_3}{42} + \frac{8}{21}\right)\beta + k_1 - 1,\\
P_4(\beta)&:=\beta^4 + 4\beta^3 + \frac{13}{3}\beta^2 - \frac{143}{48}\beta - \frac{1287}{128}.
\end{align}
\end{subequations}
The first power in \eqref{eq:EL_bulk_hedg_asympt} is dominant over the second for $\rho\to0$ if $\frac14<\beta<1$, and so $\beta$ should be chosen as a real root of $P_4$ in that interval, which however fails to exist. On the other hand, if $\beta>1$, the second power in \eqref{eq:EL_bulk_hedg_asympt} becomes dominant and $\beta$ should be chosen as a real root of $P_2$ in that range, which too fails to exist whenever $k_1>1$. Thus, \eqref{eq:beta_asympt_quartic} could be the asymptotic form of $\alpha_\lambda$ only if $\beta=1$, which makes (E) take the asymptotic form
\begin{equation}
	\label{eq:EL_bulk_hedg_asympt_beta1}
	\left(\frac{1421}{192}\pi^2\lambda^2 B^2-\frac{2717}{672}k_3-\frac{143}{32}k_1+\frac{715}{672}\right)\rho+\mathcal{O}\left(\rho^3\right)=0.
\end{equation}
Requiring the dominant power of \eqref{eq:EL_bulk_hedg_asympt_beta1} to vanish determines $B$ as in \eqref{eq:B_val}.

In a similar, but perhaps more customary way, by linearizing (E) about $\alpha=0$, we obtain that
	\begin{equation}
		\label{eq:el_rh0_to_1}
	\alpha'(\rho)\rho+2\alpha(\rho)\approx0.
	\end{equation}
By solving it subject to \eqref{eq:hom_cond}, we readily arrive at \eqref{eq:alpha_to_1} in the main text.

\section{Equivalent Dynamical System}\label{sec:eq_dyn_syst}
In this Appendix we construct a dynamical analogy for the positive branch of  equilibrium solutions $\alpha_\lambda$ for $\mathcal{F}_\lambda$ in \eqref{eq:free_energy_twisted} and give a phase space representation for them. 

We reinterpret $\mathcal{F}_\lambda$ as the action of a dynamical system by introducing the \emph{effective time}
\begin{equation}
	\label{eq:rho_to_t}
	t:=-\ln\rho.
\end{equation}
Thus, in particular,
the center of the ball $\ballx$ at $\rho=0$ is approached in the new variable when  $t\to+\infty$, while  the initial time $t=0$ corresponds to the boundary $\partial\ballx$ at $\rho=1$. Correspondingly, the twist angle $\alpha$ becomes a function on $[0,\infty)$, which is defined by 
\begin{equation}
	\label{eq:alpha_dyn}
	a(t):=\alpha\left(e^{-t}\right)
\end{equation}
and by \eqref{eq:hom_cond} satisfies
\begin{equation}
	\label{eq:a_0}
	a(0)=0.
\end{equation}
$\mathcal{F}_{\lambda}[\alpha]$ thus acquires the form of an infinite-horizon action, 
\begin{equation}
	\label{eq:free_k4_dynamic}
	\act_\lambda[a]:= \int_0^\infty \mathcal{L}_\lambda(a,\dot{a},t) \dd t,
\end{equation}
where the Lagrangian $\mathcal{L}_\lambda$ is defined as
\begin{equation}
	\label{eq:lagrangian_quartic}
	\mathcal{L}_\lambda(a,\dot{a},t):=e^{-t}\left[g(a)\dot a^2+2(k_1-1)f_0(a)\right]+\frac{32}{7}\lambda^2e^{t}\left[\sum_{n=1}^4f_n(a)\dot a^n\right]
\end{equation}
and a superimposed dot denotes differentiation with respect to $t$. 
The \emph{orbits} of the system  are  solutions of the equation of motion for $\mathcal{L}_\lambda$,
\begin{align}
	\label{eq:equation_of_motion_clcs}
	&\frac{\dd}{\dd t}\frac{\partial \mathcal{L}_\lambda}{\partial \dot a}-\frac{\partial\mathcal{L}_\lambda}{\partial a}=\nonumber\\&=e^{-t}\left[-\gamma'(a)\dot a^2-2g(a)\ddot a +2g(a)\dot a+2(k_1-1)f_0'(a)\right]\nonumber\\
	&-\frac{32}{7}\lambda^2e^{t}\left[\sum_{n=2}^4(n-1)f'_n(a)\dot a^n+\ddot a\left(\sum_{n=2}^4n(n-1)f_n(a)\dot a^{n-2}\right)+\sum_{n=1}^4nf_n(a)\dot a^{n-1}\right]=0,
\end{align}
where a prime $'$ denotes differentiation. 

We are interested in the orbits that start from the initial condition \eqref{eq:a_0}(and arbitrary $\dot{a}(0)$) and whose action $\act_\lambda$ is bounded and a minimum. To this end, we first identify the critical points of the dynamical system; these are obtained when both $\dot a\equiv0$ and $\ddot a\equiv0$ in \eqref{eq:equation_of_motion_clcs}, i.e., whenever 
\begin{equation}
	\label{eq:critical_points}
	2(k_1-1)f_0'(a)-(\lambda e^{t})^2f_1(a)=0.
\end{equation}
For $\lambda>0$, they are 
\begin{equation}
	\label{eq:crit_dyn_prob}
	a=k\pi \quad  \text{with} \quad k\in\mathbb{Z}. 
\end{equation}
For $\lambda=0$, which is the case studied in \cite{ball:brief}, they are instead
\begin{equation}
	\label{eq:crit_dyn_prob_lambda=0}
	a=k\pi \quad  \text{and} \quad a=\pm\arccos(-1/4)+2k\pi\quad\text{with}\quad k\in\mathbb{Z}. 
\end{equation}

The trajectory $a(t)\equiv 0$ represents the radial hedgehog with action $\act_\lambda[0]=0$. On the other hand,  if for $\lambda>0$ there is a trajectory $a_\lambda=a_\lambda(t)$ such that $\lim_{t\to\infty} a_\lambda(t)=\pi$ and the  action $\act_\lambda[a_\lambda]$ is finite; it remains to be seen whether $\act_\lambda[a_\lambda]<0$, to decide whether the orbit $a_\lambda$ minimizes the action. Multiplying both sides of \eqref{eq:equation_of_motion_clcs} by $\dot a$, we get
\begin{equation}
	\label{eq:equation_of_motion_clcs2}
	\begin{split}
	2g(a)\dot a e^{-t}-&\frac{32}{7}\lambda^2 e^{t}\sum_{n=1}^4nf_n(a)\dot a^n=\\
	&=e^{t}\frac{\dd}{\dd t}\left[g(a)\dot a^2-2(k_1-1)f_0(a)\right]+\frac{32}{7}\lambda^2e^{t}\frac{\dd}{\dd t}\left[\sum_{n=2}^4(n-1)f_n(a)\dot a^n\right],
	\end{split}
\end{equation}
whose integration with respect to $t\in[0,\infty)$ gives the following expression for the action of $a_\lambda$:
\begin{equation}
	\label{eq:free_energy_dyn_eq}
	\act_\lambda[a_\lambda]=-g(0)\dot{a}_\lambda(0)^2+\frac{32}{7}\lambda^2\left[\lim_{t\to\infty}\left(e^t\sum_{n=2}^4(n-1)f_n(a_\lambda)\dot a_\lambda^n\right)+2\int_0^\infty e^t\left(\sum_{n=1}^4f_n\dot a_\lambda^n\right)\dd t\right],
\end{equation}
under the assumption that the limit exists.

We are interested in \emph{bounded} orbits $a_\lambda$ with \emph{bounded} action $\act_\lambda[a_\lambda]$. We call these orbits \emph{admissible}. For a solution $a_\lambda$ of \eqref{eq:equation_of_motion_clcs} to be an admissible orbit, the initial value $\dot{a}_\lambda(0)$ must be chosen so as to ensure convergence of the orbit $a_\lambda(t)$ to $\pi$ as $t\to\infty$.

The asymptotic behaviour in \eqref{eq:alpha_to_1} here translates into
\begin{equation}
	\label{eq:t_0_dyn}
	\dot a_\lambda(t)=a_\lambda(t)+C\quad\text{for}\quad t\approx0,
\end{equation}
where $C>0$ is a constant to be determined.
One can show that, for $\lambda>0$ and material constants $k_3$ and $k_1$ chosen in the pink region of Fig.~\ref{fig:graph_sum}, there is a positive $C$ for which an admissible orbit $a_\lambda$ exists, but it has positive action $\act_\lambda$. Thus, a twisted hedgehog $\nT$ exists, but it has more energy than the radial hedgehog $\nR$, which is locally stable.
Hereafter we assume that \eqref{eq:instability_hedg} holds, so that material constants are chosen in the blue region of Fig.~\ref{fig:graph_sum}.

In the phase plane $(x,y)$, where 
\begin{equation}
	\label{eq:coordinate_phase_space}
	x(t)=a(t), \quad y(t)=\dot a(t),
\end{equation}
\eqref{eq:equation_of_motion_clcs} can be rewritten as
\begin{subequations}\label{eq:phase_plane}
\begin{align}
		&\dot x=y,\\
		&\dot y=\nonumber\\
		&\!=\!
		\frac{\left[-g'(x)y^2+2g(x)y+2(k_1-1)f_0'(x)-\frac{32}{7}(\lambda e^{t})^2\left(\sum_{n=2}^4(n-1)f'_n(x)y^{n}+\sum_{n=1}^4nf_n(x)y^{n-1}\right)\right]}{\left[2g(x)+\frac{32}{7}(\lambda e^{t})^2\sum_{n=2}^4n(n-1)f_n(x)y^{n-2}\right]},
\end{align}
\end{subequations}
a system which we next study in some detail.

\subsection{Asymptotically Autonomous Limit}\label{sec:asymptotic_autonomous}
The two-dimensional dynamical system described by \eqref{eq:phase_plane} is \emph{not} autonomous\footnote{It is explicitly dependent on time.} and this makes it  more difficult to predict the qualitative properties of its orbits, as the standard phase plane portraits (such as those discussed, for example, Chapt.\,2 of \cite{sastry:nonlinear}) do not apply here. However, system \eqref{eq:phase_plane} has the interesting property of reducing to an autonomous system in the limit as $t\to\infty$. For this reason, it is called \emph{asymptotically autonomous}.\footnote{Asymptotically autonomous dynamical systems have an interesting literature, recalled for example in Chapt.\,17 of \cite{wiggins:introduction}.}

\begin{figure}[h!] 
	\centering
	\includegraphics[width=.6\linewidth]{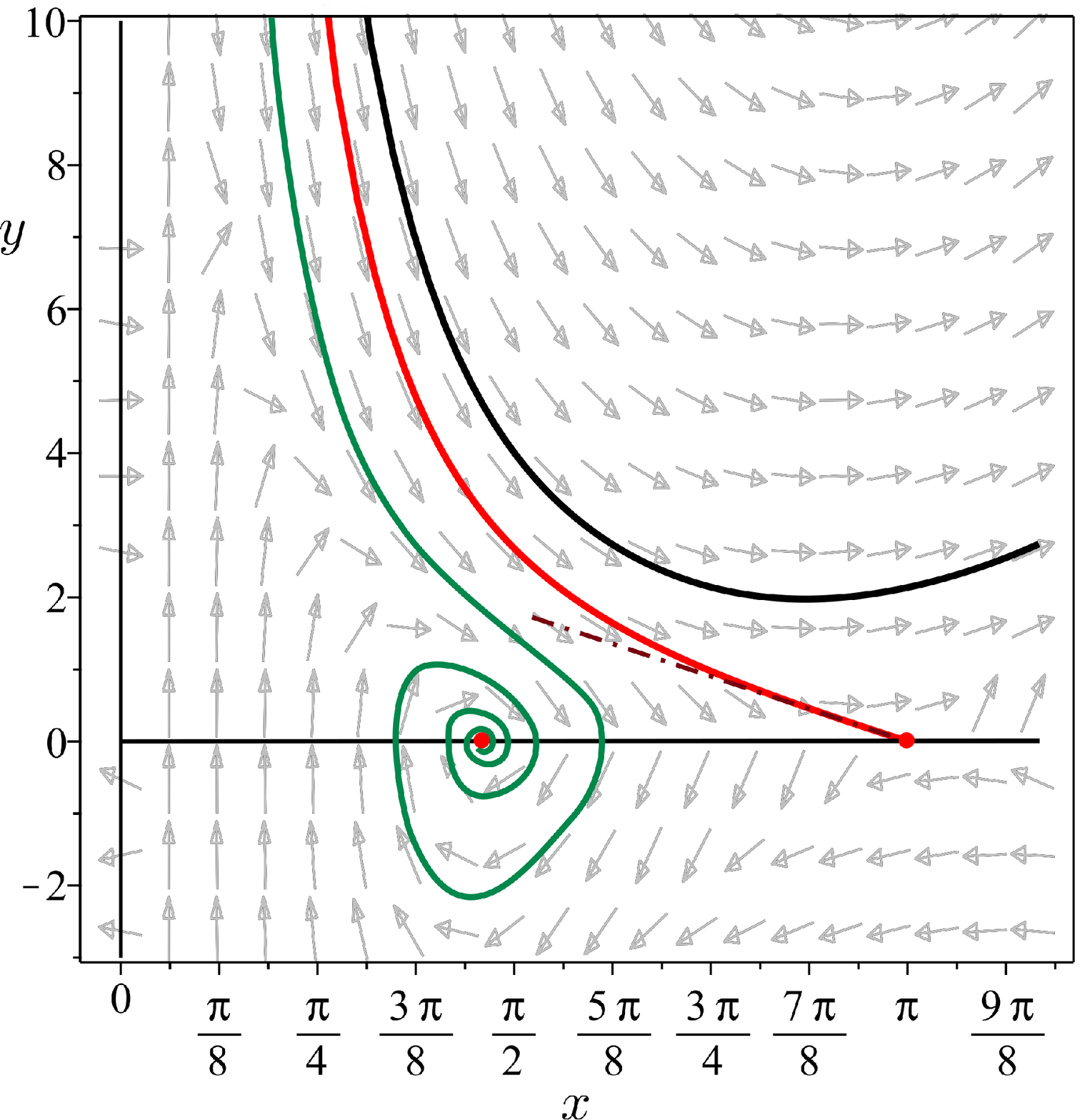}
	\caption{Phase portrait for the asymptotic autonomous limit \eqref{eq:phase_plane_t_infty} around the equilibrium points in \eqref{eq:equilibrium_asymptotic_points}, a stable spiral node and a unstable saddle. Three exemplary orbits are drawn: the green orbit spirals about the node, the blue orbit approaches the saddle along its stable invariant manifold (tangent to the broken straight line), and the black orbit, which is not connected with any equilibrium point, is unbounded.}
	\label{fig:phase_space_infty}
\end{figure}

For orbits that do not intersect either of the axes of the $(x,y)$ phase plane, the autonomous asymptotic limit of \eqref{eq:phase_plane} is
\begin{subequations}\label{eq:phase_plane_t_infty}
\begin{align}
		\dot x&=y,\\
		\dot y&=\frac{\sum_{n=2}^4(n-1)f'_n(x)y^{n}+\sum_{n=1}^4nf_n(x)y^{n-1}}{\sum_{n=2}^4n(n-1)f_n(x)y^{n-2}}.
\end{align}
\end{subequations}
For any $\lambda\geqq0$, its equilibrium points are
\begin{equation}
	\label{eq:equilibrium_asymptotic_points}
	p_1=(\arccos(1/8),0)\quad\text{and}\quad p_2=(\pi,0)
\end{equation} 
and their periodic replica. The eigenvalues of the linear approximation of \eqref{eq:phase_plane_t_infty} near these points are 
\begin{equation}
	\label{eq:eigenvalues_quartic}
	\Lambda_1^\pm=-\frac{1}{2}\pm\frac{5\sqrt{119}}{14}\mathrm{i}, \quad \Lambda_{2}^\pm=\frac{59}{44}\pm\frac{\sqrt{10015}}{44},
\end{equation}
respectively. Thus $p_1$ is a stable spiral node, while $p_2$ is a saddle. A phase portrait for the asymptotic limit \eqref{eq:phase_plane_t_infty} is shown in Fig.~\ref{fig:phase_space_infty} along with the equilibrium points in \eqref{eq:equilibrium_asymptotic_points}.\footnote{See, for example, Sect.\,2.2.2 of \cite{sastry:nonlinear}.} 
\begin{figure}[h!]
	\centering 
	\includegraphics[width=.6\linewidth]{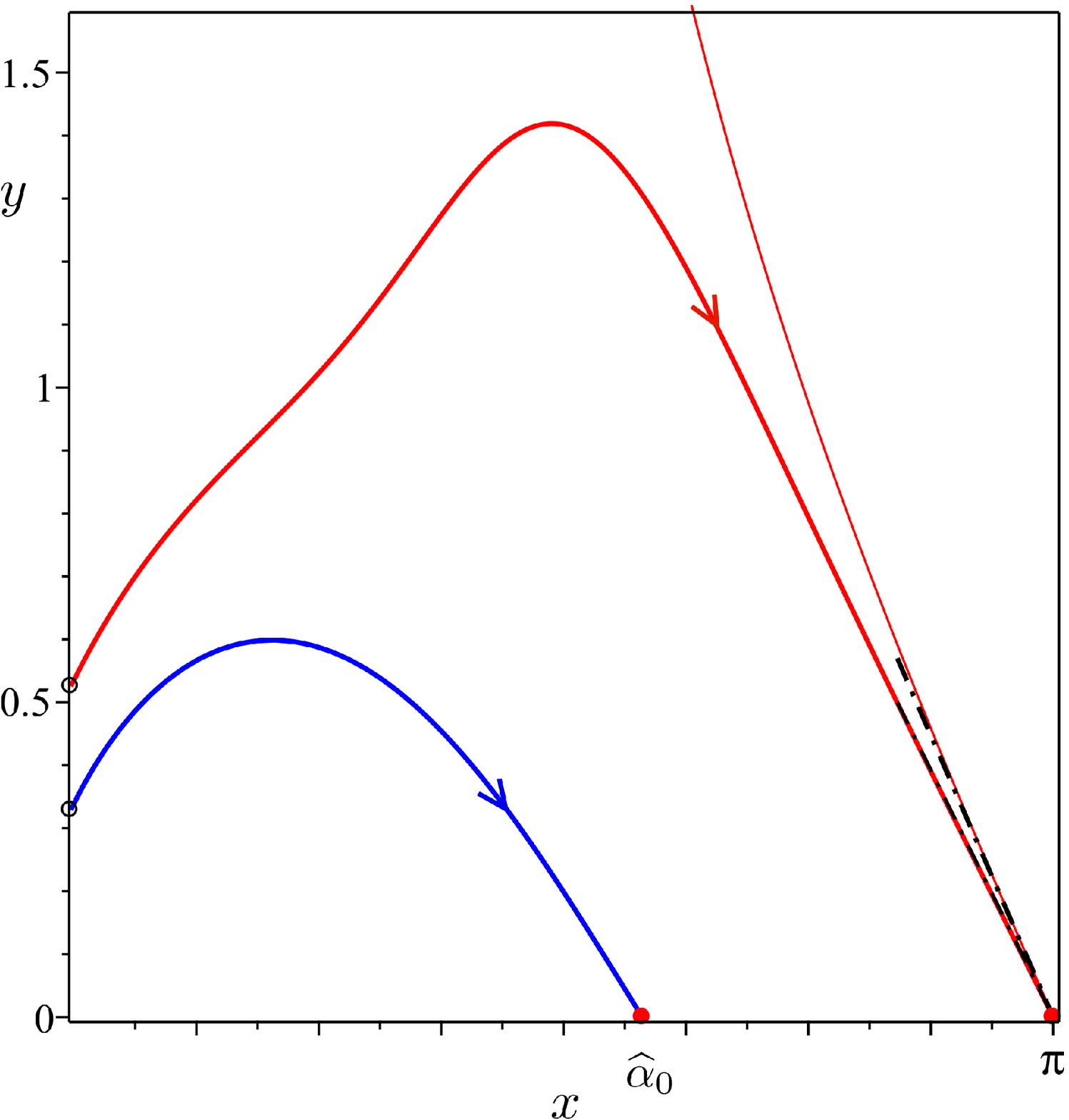}
	\caption{Admissible orbits of \eqref{eq:phase_plane} in phase space for $\lambda=0.16$ (red line) and $\lambda=0$ (blue line). The former approaches in infinite time the equilibrium point $p_2$ along the stable invariant manifold (tangent to the broken straight line), which differs from the stable invariant manifold of $p_2$ for the asymptotic autonomous limit system \eqref{eq:phase_plane_t_infty} reproduced here from Fig.~\ref{fig:phase_space_infty} (red thin line). Both orbits start from points $(0,y_0)$ on the $y$-axis; $y_0\doteq0.34$ for the blue orbit and $y_0\doteq0.54$ for the red one.}
	\label{fig:successuful_outcome}
\end{figure}

The correspondence between the solutions to an asymptotically autonomous system and those to its autonomous asymptotic limit is a delicate one and has not been completely characterized, even in the two-dimensional case, for which a larger number of results are available (see, for example, \cite{markus:asymptotically,thieme:convergence,thieme:asymptotically}). In particular, a result of Markus \cite{markus:asymptotically} (see his Theorem~7) applies to our system: it says that the $\omega$-limit set\footnote{The $\omega$-limit set of a forward solution to a dynamical system is the collection of all limiting points attained by the solution on any diverging time sequence (see, p.\,242 of \cite{wiggins:introduction} for a formal definition.)} of a solution to \eqref{eq:phase_plane} either contains the equilibria of \eqref{eq:phase_plane_t_infty} or is the union of periodic orbits of \eqref{eq:phase_plane_t_infty}.

Since \eqref{eq:phase_plane_t_infty} has no periodic orbits, we conclude that the $\omega$-limit set of any bounded solution of \eqref{eq:phase_plane} must contain the equilibria of \eqref{eq:phase_plane_t_infty}. Among these latter, only $p_2$ can be reached by an admissible orbit (according to our definition); this justifies our numerical search for a trajectory in phase space starting from a point $(0,y_0)$ of the $y$-axis and approaching in infinite time the equilibrium point $p_2$. The successful outcome of this search is shown in Fig.~\ref{fig:successuful_outcome}, where an admissible orbit of \eqref{eq:phase_plane} for $\lambda>0$ is contrasted against that obtained in \cite{ball:brief}  for $\lambda=0$, which is when \eqref{eq:phase_plane} becomes autonomous.
The solutions illustrated in Fig.~\ref{fig:successuful_outcome} are the same as those in Fig.~\ref{fig:alpha_profiles} in the main text; actually, the latter were generated from the former by inverting the change of variables in \eqref{eq:rho_to_t} and \eqref{eq:alpha_dyn}.


\end{document}